\documentclass[floatfix,aps,amsmath,nofootinbib,onecolumn,10pt]{revtex4}

\usepackage{graphicx}
\usepackage{bm}
\usepackage{rotating}
\usepackage{array}
\usepackage{makecell}
\usepackage[compatibility=false]{caption}
\usepackage{subcaption}
\def\({\left(}
\def\){\right)}
\def\[{\left[}
\def\]{\right]}
\def\e{\begin{equation}}
\def\q{\end{equation}}
\def\m{\begin{eqnarray}}
\def\n{\end{eqnarray}}

\usepackage{hyperref}

\begin{document}

\title{Constraints of dynamical dark energy models from different observational datasets}

\author{{Peiyuan Xu$^{1}$}, Lu Chen$^{1}$\footnote{Corresponding Author: Lu Chen; E-mail: chenlu@sdnu.edu.cn}, Guohao Li$^{1}$, Yang Han$^{1}$} 

\affiliation{$^1$ School of Physics and Electronics, Shandong Normal University, Jinan 250014, China}
\date{\today}

\begin{abstract}
The measurements of baryon acoustic oscillation by the Dark Energy Spectroscopic Instrument Data Release 2 indicate that dark energy may be dynamical with a time-varying equation of state. This has challenged the core assumptions of the $\Lambda$CDM model and aroused widespread discussion. Existing work has achieved fruitful results in the dark energy models, exploring various parameterization forms, but it lacks systematic parameter constraints based on the latest dataset combinations. We use $\Lambda$CDM as a baseline model and carry out rigorous statistical constraints on key cosmological parameters for seven representative parameterization models. The Planck PR4 and DESI DR2 observations are incorporated into our study. We use four datasets: CMB+BAO+PantheonPlus, CMB+BAO+DES-Y5, CMB+BAO+Union3, and CMB+BAO(without LRG1 and LRG2)+DES-Y5. Our results may not effectively alleviate ${H}_{0}$ tension, but may relatively reduce ${\sigma }_{8}$ tension. By comparing the Akaike Information Criterion and the Bayesian evidence obtained for each model, we demonstrate that the linear Chevallier-Polarski-Linder parameterization is not the optimal choice in all cases. The Logarithmic model shows the best fitting performance among three different SNIa samples. However, with CMB+BAO(without LRG1 and LRG2)+DES-Y5, Jassal-Bagla-Padmanabhan and Chevallier-Polarski-Linder models gain more obvious preference. All two-parameter dynamical dark energy models perform most prominently in the CMB+BAO+DES-Y5 dataset, with the Logarithmic model providing strong evidence to support dynamical dark energy.


\end{abstract}

\pacs{???}

\maketitle


\section{Introduction} 
\label{sec:int}
In 1998, observations of distant Type Ia supernovae (SNIa) first suggested that our universe is undergoing accelerated expansion, which is one of the most important cosmological discoveries in the past three decades\cite{SupernovaSearchTeam:1998fmf,SupernovaCosmologyProject:1998vns}. The most successful theory to explain this phenomenon is dark energy. In the standard model of cosmology, the $\Lambda$-Cold Dark Matter ($\Lambda$CDM) model \cite{Bamba:2012cp}, the universe is composed of cold dark matter, baryons, negligible radiation and dark energy, where dark energy is considered to be the cosmological constant $\Lambda$, accounting for about 70\% of the energy in the universe. Although the $\Lambda$CDM model has been in excellent agreement with a variety of observations for a long time, it also faces challenges from some well-known theoretical problems, such as Hubble tension, $\sigma_8$ tension et. al. \cite{Verde:2019ivm, DiValentino:2021izs, Riess:2021jrx, Wu:2024faw, RoyChoudhury:2024wri, RoyChoudhury:2025dhe, RoyChoudhury:2025iis, Chaudhary:2025pcc, Chaudhary:2025uzr, Chaudhary:2025bfs, Nagpal:2025omq, Silva:2025twg, Das:2023rvg, Das:2013sca}. 

Hubble tension refers to the significant difference between the value of the Hubble constant ${H}_{0}$ inferred from the Cosmic Microwave Background (CMB) and the value obtained through observations of the local distance ladder. In the $\Lambda$CDM framework, the value obtained from the Planck CMB+lensing is: $67.36 \pm 0.54$ $\mathrm{km/s/Mpc}$ \cite{Planck:2018vyg}. However, local measurements conducted by the SH0ES team using Cepheid variable calibrators show ${H}_{0}$ = $73.04 \pm 1.04$ $\mathrm{km/s/Mpc}$ \cite{Riess:2021jrx}, a difference of 4.85$\sigma$. In comparison, the ${\sigma }_{8}$ tension is relatively low. The values of ${\sigma }_{8}$ obtained from Weak Lensing (WL), Cluster Counts (CC), or Redshift Space Distortion (RSD) are all significantly lower than the value of $0.811 \pm 0.006$ given by Planck CMB TTTEEE+lowE+lensing \cite{Planck:2018vyg}, with a difference of 2-3$\sigma$. These tensions pose increasing challenges to the standard cosmological model and may suggest the emergence of new physics beyond the $\Lambda$CDM model.

\par To better address these issues, the parameterized Dynamical Dark Energy (DDE) model emerges. Typically, the Dark Energy Equation of State (DE EOS) can be defined as $w=\frac{p_{\text{de}} }{\rho _{\text{de}} }$ to characterize the properties of dark energy, where $p_{\text{de}}$ represents the pressure of dark energy and $\rho _{\text{de}}$ represents its energy density. The DDE model assumes that $w$ is a constant different from -1 or evolves with time, thus constructing a flexible theoretical framework that allows for a more systematic exploration of potential deviations from the cosmological constant $\Lambda$. Various parameterized DDE models have been proposed. Among them, the $w$CDM model is a minimal extension of the $\Lambda$CDM model, in which the DE EOS $w$ is kept constant other than -1. The most representative and widely accepted two-parameter extension model is the CPL model. It can be expressed as: $w(a)={w}_{0}+{w}_{a}(1-a)$.

\par From an observational perspective, $Planck$ investigation of the anisotropy for CMB is of great significance to cosmology. However, the CMB data alone impose relatively limited constraints on the DDE model. Even in simple two-parameter models designed to minimize degrees of freedom, the constraints generated by CMB are often too broad to provide rich and useful information. Therefore, observations of the local universe have assumed a more prominent position.

\par SNIa DES-Y5 \cite{DES:2024jxu, DES:2024upw, DES:2024hip}, based on a joint compilation of PantheonPlus \cite{Scolnic:2021amr, Brout:2022vxf} and Union3 \cite{Rubin:2023jdq}, find that, whether using SNIa alone or in combination with CMB and Baryon Acoustic Oscillations (BAO), the best fit EOS $w$ is always slightly greater than -1 above the 1$\sigma$ level. This is consistent with the constraints of Union3 and supports the trend of DDE. 
The Dark Energy Spectroscopic Instrument (DESI) \cite{Sousa-Neto:2025gpj, DESI:2024mwx, DESI:2024aqx, DESI:2025zgx, DESI:2024kob} tracks the expansion of the universe using BAO observations. When DESI DR1 is combined with Planck CMB and SNIa (PantheonPlus, DES-Y5 and Union3), the results show a clear tension with the $\Lambda$CDM model, giving a strong indication of DDE. Specifically, in the CPL model, ${w}_{0}$ $>$ -1 and ${w}_{a}$ $<$ 0 have a clear preference at statistical levels between 2.5$\sigma$ and 3.9$\sigma$ \cite{DESI:2024mwx}. 
The newly released DESI DR2 data further strengthen this preference \cite{DESI:2025zgx, DESI:2025fii, DESI:2025qqy}. The combination of DESI DR2 with CMB alone yields a significance of 3.1$\sigma$ for $w$ to the exclusion of $\Lambda$CDM. When combined with CMB, DESI DR2 and SNIa, this preference for DDE still exists, reaching significances of 2.8$\sigma$, 3.8$\sigma$ and 4.2$\sigma$, respectively \cite{DESI:2025zgx}. This has sparked an extensive discussion about dark energy. 
Inconsistencies between different datasets may be a reason for these discrepancies. For example, excluding LRG1 (${z}_{\text{eff}}$ = 0.51) and LRG2 (${z}_{\text{eff}}$ = 0.71) individually results in $w$ = -1 falling within the 2$\sigma$ contour \cite{Wang:2024pui, Liu:2024gfy}. In addition, the BAO measurements of the Sloan Digital Sky Survey (SDSS) and DESI collaboration are also inconsistent with each other \cite{Ghosh:2024kyd, Wolf:2025jlc, Afroz:2025iwo, Cheng:2025lod, Fazzari:2025lzd, Lee:2022cyh, Colgain:2025nzf, Colgain:2024xqj, Vagnozzi:2023nrq, Pedrotti:2025ccw}.

\par To comprehensively analyze various DDE parameterizations, we integrate cutting-edge observational data from current cosmological research, focusing on the recently released CMB PR4 and DESI DR2 measurements. Based on these updated datasets and combined with three representative SNIa samples, we construct four core dataset combinations: CMB+BAO+PantheonPlus, CMB+BAO+DES-Y5, CMB+BAO+Union3 and CMB+BAO(without LRG1 and LRG2)+DES-Y5. Using the $\Lambda$CDM model as a benchmark framework, we conduct a comprehensive parameter constraint analysis on seven representative DDE parameterizations: $w$CDM, Chevallier-Polarski-Linder (CPL), Jassal-Bagla-Padmanabhan (JBP), Feng-Shen-Li-Li (FSLL), Barboza-Alcaniz (BA), Logarithmic (LOG) and Exponential (EXP).

This paper is organized as follows. 
In Section~\ref{sec:ion}, we introduce the Friedmann equations and the evolution of dark energy. 
Section~\ref{sec:res} describes the methods employed in our analysis and all observational datasets used.
In Section~\ref{results}, we give constraints on the relevant parameters for all DDE parameterized models.
Section~\ref{sec:sum} presents a comparative analysis of models and observational data, systematically discussing our results.
Finally, a brief summary is included in Section~\ref{summary}.

\section{Dynamic Dark Energy Models}
\label{sec:ion}
The Friedmann equation describes the evolution of our universe. Considering a statistically homogeneous and isotropic universe that satisfies the Friedmann-Lemaître-Robertson-Walker (FLRW) metric, we can obtain:\\
\begin{equation}
\begin{aligned}
{H}^{2}(a)=\frac{8\pi G}{3}\left[{\rho }_{\text{r},0}{a}^{-4}+{\rho }_{\text{m},0}{a}^{-3}+{\rho }_{\text{de},0}X(a)\right].\\    
\end{aligned}
\end{equation}
Here $a$ represents the scale factor, $H(a)$ denotes the Hubble parameter, and $G$ is the Newton's gravitational constant. $\rho$ is the energy density, the subscript $0$ indicates current values of the physical quantities, while the subscripts $\text{{r}}$, $\text{{m}}$, and $\text{{de}}$ are the radiation, matter, and dark energy, respectively. Meanwhile,
\begin{equation}
\begin{aligned}
X(a)=\text{Exp}\left[-3\int_{{a}_{0}=1}^{a}\frac{1+w(a)}{a}da\right].\\    
\end{aligned}
\end{equation}
Where $w(a)$ represents the DE EOS as a function of the scale factor and can be parameterized by ${w}_{0}$ and ${w}_{a}$. 
In section~\ref{results}, we will list the $X(a)$ of each DDE model separately to illustrate the evolution of dark energy density more intuitively.\\

\section{Methods}
\label{sec:res} 
In this section, we summarize the statistical methods employed in our analysis and list the combination of observational data used.\\
\subsection{Statistical Analyses}
We use the publicly available cosmological calculation code CAMB \cite{Lewis:1999bs, Howlett:2012mh} and make specific modifications for different models. We select the advanced nested sampling algorithm PolyChord \cite{Handley:2015fda, Handley:2015vkr} for parameter space exploration and analysis with the publicly available sampler Cobaya \cite{Lewis:2002ah, Lewis:2013hha}. Based on slice sampling, PolyChord is particularly suited for exploring higher dimensional parameter spaces. We set ``precision\_criterion: 0.001'' as the standard for program stopping. In addition, this algorithm can accurately compute the Bayesian evidence $\text{ln}\mathcal{Z}$ of each model while also generating posterior samples. $\text{ln}\mathcal{Z}$ quantifies how much the data prefer a given model. The smaller the absolute value, the stronger the support for the model. We adopt $\Lambda$CDM as a benchmark and compare models using $\Delta\text{ln}\mathcal{Z}$: if $\Delta\text{ln}\mathcal{Z} < 0$, the standard model is more favorable; if $\Delta\text{ln}\mathcal{Z} > 0$, the data prefer the alternative model. The Jeffreys scale provides a quantitative assessment of this preference \cite{Jeffreys:1939xee, Kass:1995loi}. Generally, $0\leq\left| \Delta\text{ln}\mathcal{Z}\right|<1$ is considered weak evidence; $1\leq\left|\Delta\text{ln}\mathcal{Z}\right|<3$ is definite or moderate evidence; $3\leq\left| \Delta\text{ln}\mathcal{Z}\right|<5$ represents strong evidence and $\left|\Delta\text{ln}\mathcal{Z}\right|\geq5$ indicates decisive evidence.


All DDE models listed in Section~\ref{results} can be characterized by 8 free parameters: \{${\Omega }_{\text{c}}{h}^{2}$, ${\Omega }_{\text{b}}{h}^{2}$, ${A}_{\text{s}}$, ${n}_{\text{s}}$, $\tau$, ${\theta }_{\text{MC}}$, ${w}_{\text{0}}$, ${w}_{a}$\}. ${\Omega }_{\text{c}}{h}^{2}$ is the physical cold dark matter energy density; ${\Omega }_{\text{b}}{h}^{2}$ is the physical baryon energy density; ${A}_{\text{s}}$ is the primordial scalar spectrum amplitude; ${n}_{\text{s}}$ is the spectrum index; $\tau$ is the reionization optical depth; ${\theta }_{\text{MC}}$ is the angular size of the sound horizon. The DE EOS has two free parameters: ${w}_{\text{0}}$, which is its current value, and ${w}_{a}$, which describes its time variation. In TABLE~\ref{tab:parameters}, we give the flat prior ranges of the free variation of the parameters. 

\begin{table}[h]
\centering
\begin{tabular}{l l}
\hline \hline
Parameter & Prior \\
\hline
$\Omega_{\mathrm{b}} h^2$ & $[0.005, 0.1]$ \\
$\Omega_{\mathrm{c}} h^2$ & $[0.001, 0.99]$ \\
$\log(10^{10} A_{\mathrm{s}})$ & $[1.61, 3.91]$ \\
$n_{\mathrm{s}}$ & $[0.8, 1.2]$ \\
$\tau$ & $[0.01, 0.8]$ \\
$100\theta_{\mathrm{MC}}$ & $[0.5, 10]$ \\
$w_0$ & $[-3, 1]$ \\
$w_a$ & $[-6, 2]$ \\
\hline \hline
\end{tabular}
\caption{The range of flat prior distributions of free cosmological parameters in our analysis.}
\label{tab:parameters}
\end{table}

\subsection{Datasets}
Here we list all data and the corresponding references used in our analysis.
\begin{itemize}

\item \textbf{CMB:} We use Planck 2018 PR3 low-$l$ TT Commander likelihood in the range of $2 \leq l \leq 29$ and Planck 2018 PR3 low-$l$ SimAll EE likelihood at $2 \leq l \leq 29$ \cite{Planck:2018vyg}. In addition, we employ the latest Planck PR4 CamSpec high-$l$ temperature and polarization likelihood based on NPIPE maps, covering TTTEEE power spectra from $l$ = 30 to $l$ = 2500 \cite{Rosenberg:2022sdy, Efstathiou:2019mdh}. Furthermore, we incorporate CMB lensing measurements, which are reconstructed from the CMB lensing potential using Planck PR4 NPIPE maps \cite{Carron:2022eyg}.


\item \textbf{BAO:}
We adopt the BAO measurements from DESI DR2 in the range of $0.1 < z < 4.2$ \cite{DESI:2025zgx, DESI:2025zpo}. These measurements cover a wide range of redshifts and are obtained from various tracers, including the Bright Galaxy Sample (BGS), Luminous Red Galaxies (LRG), Emission Line Galaxies (ELG), Quasars (QSO), and the Lyman-$\alpha$ (Ly$\alpha$) forests. 
DESI DR2 provides seven measurements at the effective redshift ($z_{\text{eff}}$). To be specific, BGS: $z_{\text{eff}}$ = 0.295, LRG1: $z_{\text{eff}}$ = 0.510, LRG2: $z_{\text{eff}}$ = 0.706, LRG3+ELG1: $z_{\text{eff}}$ = 0.934, ELG2: $z_{\text{eff}}$ = 1.321, QSO: $z_{\text{eff}}$ = 1.484, Ly$\alpha$: $z_{\text{eff}}$ = 2.330.

LRG1 ($z_{\text{eff}}$ = 0.510) and LRG2 ($z_{\text{eff}}$ = 0.706) in DESI DR2 may have an impact on its conclusions regarding DDE evidence, which is controversial. Therefore, we remove LRG1 and LRG2 from DESI DR2 and compare them with other datasets containing them for further exploration.

\item \textbf{SNIa:} 
As standard candles, the SNIa determine the cosmic distances by using the cosmic distance ladder. They are powerful cosmological probes. For this analysis, we use three different SNIa data to improve the constraints on the cosmological parameters. 
Firstly, we utilize the PantheonPlus sample \cite{Scolnic:2021amr, Brout:2022vxf} consisting of 1550 SNIa over the redshift range $0.001 < z < 2.26$. Then we also employ the DES-Y5 sample \cite{DES:2024jxu, DES:2024upw, DES:2024hip}, which includes 1635 light curves from 1550 SNIa in the redshift range $0.10 < z < 1.13$, along with 194 low-redshift SNIa in the range $0.025 < z < 0.10$.
For PantheonPlus and DES-Y5, we use the full luminosity distance likelihood (non-compressed data), including the full covariance matrix. The light curve in the analysis is processed using the SALT3 model \cite{Kenworthy:2021azy}. Lastly, we use an updated “Union” compilation consisting of 2087 cosmologically useful SNIa from 24 datasets (“Union3”) \cite{Rubin:2023jdq}. It is worth noting that 1363 of these SNIa overlap with the PantheonPlus. This dataset adopts Bayesian Hierarchical Modeling, which uniquely handles systematic errors and uncertainties \cite{Scherer:2025esj}.
Unlike the first two types of SNIa data, Union3 uses the compressed distance modulus likelihood and does not include the light curve parameters. Union3 is calibrated according to the SALT framework \cite{SNLS:2007cqk, SupernovaCosmologyProject:2011ycw, Kenworthy:2021azy} optical curve fitting.

\end{itemize}

We test these models using four different datasets: CMB+BAO+PantheonPlus, CMB+BAO+Union3, CMB+BAO+DES-Y5 and CMB+BAO(without LRG1 and LRG2)+DES-Y5. It should be noted that PantheonPlus and Union3 share 1363 overlapping SNIa data. For this reason, we combine the relatively independent DES-Y5 with BAO (without LRG1 and LRG2). We then incorporate CMB to construct a new dataset: CMB+BAO(without LRG1 and LRG2)+DES-Y5. It allows a straightforward comparison with the constraints from CMB+BAO+DES-Y5.

\section{Results}
\label{results}
In this section, we detail the constraints on the $\Lambda$CDM and seven DDE models included in our study. Subsequently, we analyze the parameter results of each model in turn. The results of all tests are presented in Table~\ref{tb:result_combined}. This table combines horizontal and vertical presentation: horizontally, the parameter results for the same model under different datasets are compared, while vertically, the parameter results for all models under the same dataset are presented. 

\par The parameters discussed in this article mainly include two free parameters: ${w}_{0}$ and ${w}_{a}$; three derived parameters: ${\Omega }_{m}$, ${\sigma }_{8}$, and ${H}_{0}$. We only list the above because the changes of other parameters compared with the Planck CMB are very small and can be ignored. Figure~\ref{FIG1} shows the full triangle plots of the $\Lambda$CDM, $w$CDM, CPL, JBP, FSLL I, FSLL II, BA, LOG, and EXP models under the constraints of four datasets: CMB+BAO+PantheonPlus, CMB+BAO+DES-Y5, CMB+BAO+Union3, and CMB+BAO(without LRG1 and LRG2)+DES-Y5 respectively. Figure~\ref{FIG2} shows the complete triangle diagrams of all two-parameter DDE models under the constraints of the four datasets described above. Figure~\ref{FIG3} exhibits the evolution of $w(z)$ and ${\rho}_{de}(z)/{\rho}_{de}(0)$ as functions of the redshift $z$ for all models. Since the models show similar evolutionary trends under different datasets with no qualitative differences, we only present the results from CMB+BAO+DES-Y5 for illustration.

\subsubsection{$\Lambda$CDM}
The $\Lambda$CDM is included as the standard model in our work, where ${w}_{0}$ = -1, ${w}_{a}$ = 0, to compare with other models. The parameter constraints of $\Lambda$CDM are little affected by different SNIa data, showing high stability. When using the combination of CMB+BAO(without LRG1 and LRG2)+DES-Y5, both ${H}_{0}$ and ${\Omega }_{m}$ exhibit slight changes, with the former decreasing and the latter increasing. As can be seen in Figure~\ref{subfig:a}, ${H}_{0}$ is inversely proportional to ${\Omega }_{\text{m}}$. 

We compare our ${H}_{0}$ results (68\% C.L.) with ${H}_{0}$ = $67.36\pm0.54$ $\mathrm{km/s/Mpc}$ given by Planck PR3 TTTEEE+lowE+lensing (hereafter Planck PR3) \cite{Planck:2018vyg} and the local measurement result of ${H}_{0}$ = $73.04 \pm 1.04$ $\mathrm{km/s/Mpc}$ (hereafter Local) \cite{Riess:2021jrx}. Under CMB+BAO+PantheonPlus, ${H}_{0}$ = $68.09\pm0.25$ $\mathrm{km/s/Mpc}$, 1.23$\sigma$ difference compared to Planck PR3 and 4.63$\sigma$ difference compared to Local. Under CMB+BAO+DES-Y5, ${H}_{0}$ = $68.03\pm0.24$ $\mathrm{km/s/Mpc}$, 1.13$\sigma$ difference compared to Planck PR3 and 4.69$\sigma$ difference compared to Local. In CMB+BAO+Union3, ${H}_{0}$ = $68.12\pm0.28$ $\mathrm{km/s/Mpc}$, resulting in a 1.25$\sigma$ difference compared to Planck PR3 and a 4.57$\sigma$ difference compared to Local.
Excluding LRG1 and LRG2 from BAO in CMB+BAO+DES-Y5 leads to a further decrease in $H_0$ to $67.72_{-0.32}^{+0.27}$ $\mathrm{km/s/Mpc}$, without alleviating the $H_0$ tension.


The change of ${\sigma }_{8}$ is also not obvious. Taking CMB+BAO+Union3 as an example, we find that the value of ${\sigma }_{8}$ is $0.8054_{-0.0063}^{+0.0054}$ at 68\% C.L.. Compared to the value ${\sigma }_{8}$ = $0.811\pm0.006$ under Planck PR3, the difference is minimal, just 0.69$\sigma$. Compared to the ${\sigma }_{8}$ value of $0.819\pm0.015$ obtained from the combination of ACT DR6 and BAO (6df and SDSS) \cite{ACT:2023kun}, there is a 0.85$\sigma$ discrepancy. Compared with ${\sigma }_{8}$ = $0.783_{-0.092}^{+0.073}$ from DES-Y3 \cite{DES:2021bvc, DES:2021vln}, the difference is 0.31$\sigma$. Although the difference between our result and DES-Y3 is reduced, it is mainly because the error bar of the data provided by DES-Y3 is too large, which leads to broad constraints. Our results in $\Lambda$CDM do not provide new implications for the ${\sigma }_{8}$ tension.

{%
\setlength{\tabcolsep}{8pt}
\renewcommand{\arraystretch}{1.0}
\footnotesize

\begin{table}[htbp]
\caption{The 68\% limits of the cosmological parameters for all models under the combination of CMB+BAO+ PantheonPlus, CMB+BAO+DES-Y5, CMB+BAO+Union3 datasets and CMB+BAO(without LRG1 and LRG2) +DES-Y5.}
\label{tb:result_combined}
    \centering
    \begin{tabular}{@{}l|cccc@{}}
        \hline\hline
        Model & CMB+BAO+PantheonPlus
              & CMB+BAO+DES-Y5
              & CMB+BAO+Union3
              & \makecell{CMB+BAO(without \\LRG1 and LRG2)+DES-Y5} \\
        \hline\hline
        
        \multicolumn{5}{l}{\textbf{$\Lambda$CDM}} \\
        \hline
        $\Omega_\text{m}$ & $0.3037_{-0.0031}^{+0.0034}$ & $0.3046_{-0.0034}^{+0.0029}$ & $0.3033_{-0.0035}^{+0.0039}$ & $0.3086_{-0.0036}^{+0.0043}$ \\
        $\sigma_8$ & $0.8049_{-0.0051}^{+0.0047}$ & $0.8055\pm0.0053$ & $0.8054_{-0.0063}^{+0.0054}$ & $0.8072_{-0.0058}^{+0.0054}$ \\
        $H_0~[\mathrm{km/s/Mpc}]$ & $68.09\pm0.25$ & $68.03\pm0.24$ & $68.12\pm0.28$ & $67.72_{-0.32}^{+0.27}$ \\
        \hline
        $\chi^2$ & $12406.2$ & $12650.2$ & $11028.5$ & $12638.4$\\
        AIC & $12418.2$ & $12662.2$ & $11040.5$ & $12650.4$ \\
        $\Delta\text{ln}\mathcal{Z}$ & 0.00 & 0.00 & 0.00 & 0.00 \\
        \hline\hline

        \multicolumn{5}{l}{\textbf{$w$CDM}} \\
        \hline
        $w_0$ & $-0.993_{-0.019}^{+0.020}$ & $-0.968\pm0.018$ & $-0.996_{-0.025}^{+0.028}$ & $-0.972\pm0.021$ \\
        \hline
        $\Omega_\text{m}$ & $0.3045_{-0.0050}^{+0.0045}$ & $0.3101_{-0.0041}^{+0.0044}$ & $0.3040\pm0.0055$ & $0.3133_{-0.0050}^{+0.0059}$ \\
        $\sigma_8$ & $0.8040_{-0.0076}^{+0.0077}$ & $0.7969_{-0.0072}^{+0.0083}$ & $0.8045_{-0.0096}^{+0.0092}$ & $0.7984_{-0.0076}^{+0.0087}$ \\
        $H_0~[\mathrm{km/s/Mpc}]$ & $67.97\pm0.50$ & $67.29_{-0.47}^{+0.45}$ & $68.04\pm0.63$ & $67.10_{-0.58}^{+0.52}$ \\
        \hline
        $\chi^2$ & $12406.5$ & $12647.3$ & $11030.6$ & $12637.5$ \\
        AIC & $12420.5$ & $12661.3$ & $11044.6$ & $12651.5$\\
        $\Delta\text{ln}\mathcal{Z}$ & -3.11 & -2.05 & -4.46 & -3.69 \\
        \hline\hline

        \multicolumn{5}{l}{\textbf{CPL}} \\
        \hline
        $w_0$ & $-0.841_{-0.056}^{+0.050}$ & $-0.754_{-0.056}^{+0.053}$ & $-0.676_{-0.091}^{+0.093}$ & $-0.761_{-0.067}^{+0.060}$\\
        $w_a$ & $-0.594_{-0.207}^{+0.200}$ & $-0.844_{-0.201}^{+0.234}$ & $-1.045_{-0.283}^{+0.313}$ & $-0.863_{-0.239}^{+0.285}$\\
        \hline
        $\Omega_\text{m}$ & $0.3109_{-0.0050}^{+0.0054}$ & $0.3186_{-0.0051}^{+0.0056}$ & $0.3263_{-0.0087}^{+0.0088}$ & $0.3173_{-0.0057}^{+0.0053}$\\
        $\sigma_8$ & $0.8078_{-0.0077}^{+0.0083}$ & $0.8029_{-0.0079}^{+0.0084}$ & $0.7964_{-0.0092}^{+0.0090}$ & $0.8050\pm0.0077$\\
        $H_0~[\mathrm{km/s/Mpc}]$ & $67.52_{-0.56}^{+0.52}$ & $66.75_{-0.57}^{+0.53}$ & $65.98_{-0.95}^{+0.70}$ & $66.94_{-0.53}^{+0.55}$\\
        \hline
        $\chi^2$ & $12398.8$ & $12632.4$ & $11015.2$ & $12624.3$\\
        AIC & $12414.8$ & $12648.4$ & $11031.2$ & $12640.3$\\
        $\Delta\text{ln}\mathcal{Z}$ & -1.02 & 4.45 & 2.16 & 2.00 \\
        \hline\hline

        \multicolumn{5}{l}{\textbf{JBP}} \\
        \hline
        ${w}_{0}$ & $-0.813_{-0.080}^{+0.072}$ & $-0.663_{-0.086}^{+0.078}$ & $-0.542_{-0.129}^{+0.115}$ & $-0.661_{-0.080}^{+0.092}$\\ 
        ${w}_{a}$ & $-1.118_{-0.427}^{+0.456}$ & $-1.851_{-0.446}^{+0.520}$ & $-2.421_{-0.613}^{+0.685}$ & $-1.967_{-0.555}^{+0.488}$\\
        \hline
        ${\Omega }_{m}$ & $0.3098_{-0.0060}^{+0.0058}$ & $0.3193_{-0.0061}^{+0.0055}$ & $0.3286_{-0.0089}^{+0.0087}$ & $0.3187\pm0.0056$\\
        ${\sigma }_{8}$ & $0.8049_{-0.0088}^{+0.0091}$ & $0.7977_{-0.0089}^{+0.0083}$ & $0.7902_{-0.0100}^{+0.0096}$ & $0.8015\pm0.0080$\\
        ${H}_{0}~[\mathrm{km/s/Mpc}]$ & $67.56\pm0.63$ & $66.56\pm0.59$ & $65.66_{-0.88}^{+0.84}$ & $66.74\pm0.55$\\
        \hline
        ${\chi }^{2}$ & $12402.6$ & $12635.4$ & $11017.2$ & $12624.5$\\
        AIC & $12418.6$ & $12651.4$ & $11033.2$ & $12640.5$\\
        $\Delta\text{ln}\mathcal{Z}$ & -4.33 & 1.52 & -1.65 & 2.76\\
        \hline\hline

        \multicolumn{5}{l}{\textbf{FSLL I}} \\
        \hline
        ${w}_{0}$ & $-0.844\pm0.060$ & $-0.741_{-0.063}^{+0.058}$ & $-0.643\pm0.099$ & $-0.748_{-0.070}^{+0.065}$\\
        ${w}_{a}$ & $-0.543_{-0.209}^{+0.197}$ & $-0.831_{-0.213}^{+0.224}$ & $-1.089_{-0.303}^{+0.305}$ & $-0.842_{-0.241}^{+0.271}$\\
        \hline
        ${\Omega }_{m}$ & $0.3102_{-0.0058}^{+0.0057}$ & $0.3177_{-0.0051}^{+0.0055}$ & $0.3266_{-0.0093}^{+0.0085}$ & $0.3178\pm0.0056$\\
        ${\sigma }_{8}$ & $0.8052_{-0.0088}^{+0.0083}$ & $0.7991_{-0.0083}^{+0.0081}$ & $0.7922_{-0.0090}^{+0.0102}$ & $0.8025_{-0.0089}^{+0.0081}$\\
        ${H}_{0}~[\mathrm{km/s/Mpc}]$ & $67.53_{-0.64}^{+0.57}$ & $66.73_{-0.54}^{+0.51}$ & $65.86_{-0.87}^{+0.84}$ & $66.83_{-0.58}^{+0.56}$\\
        \hline
        ${\chi }^{2}$ & $12401.0$ & $12634.2$ & $11018.1$ & $12626.1$\\
        AIC & $12417.0$ & $12650.2$ & $11034.1$ & $12642.1$\\
        $\Delta\text{ln}\mathcal{Z}$ & -3.46 & 3.22 & -0.53 & 0.03 \\
        \hline\hline

        \multicolumn{5}{l}{\textbf{FSLL II}} \\
        \hline
        ${w}_{0}$ & $-0.898_{-0.036}^{+0.034}$ & $-0.857\pm0.034$ & $-0.814\pm0.052$ & $-0.860_{-0.044}^{+0.034}$\\
        ${w}_{a}$ & $-0.565_{-0.177}^{+0.193}$ & $-0.704_{-0.183}^{+0.208}$ & $-0.842_{-0.218}^{+0.273}$ & $-0.724_{-0.199}^{+0.291}$\\
        \hline
        ${\Omega }_{m}$ & $0.3113_{-0.0052}^{+0.0055}$ & $0.3167_{-0.0051}^{+0.0049}$ & $0.3227_{-0.0076}^{+0.0078}$ & $0.3159\pm0.0056$\\
        ${\sigma }_{8}$ & $0.8092\pm0.0086$ & $0.8052\pm0.0080$ & $0.8006\pm0.0098$ & $0.8069\pm0.0085$\\
        ${H}_{0}~[\mathrm{km/s/Mpc}]$ & $67.54_{-0.60}^{+0.55}$ & $66.98\pm0.50$ & $66.38_{-0.79}^{+0.76}$ & $67.09\pm0.56$\\
        \hline
        ${\chi }^{2}$ & $12396.4$ & $12631.5$ & $11015.6$ & $12626.4$\\
        AIC & $12412.4$ & $12647.5$ & $11031.6$ & $12642.4$\\
        $\Delta\text{ln}\mathcal{Z}$ & -1.42 & 3.55 & 2.23 & -0.46 \\
        \hline\hline
    \end{tabular}
\end{table}

\newpage
{%
\setlength{\tabcolsep}{8pt}
\renewcommand{\arraystretch}{1.0}
\footnotesize

\begin{table}[htbp]
\caption*{\tablename{} \ref{tb:result_combined} (Continued)}
    \centering
    \begin{tabular}{@{}l|cccc@{}}
        \hline\hline
        Model & CMB+BAO+PantheonPlus
              & CMB+BAO+DES-Y5
              & CMB+BAO+Union3
              & \makecell{CMB+BAO(without \\LRG1 and LRG2)+DES-Y5} \\
        \hline\hline

        \multicolumn{5}{l}{\textbf{BA}} \\
        \hline
        ${w}_{0}$ & $-0.865\pm0.046$ & $-0.791\pm0.044$ & $-0.730_{-0.073}^{+0.077}$ & $-0.804\pm0.051$\\
        ${w}_{a}$ & $-0.290_{-0.089}^{+0.111}$ & $-0.396_{-0.090}^{+0.103}$ & $-0.483\pm0.127$ & $-0.402_{-0.105}^{+0.120}$\\
        \hline
        ${\Omega }_{m}$ & $0.3110_{-0.0060}^{+0.0058}$ & $0.3191\pm0.0052$ & $0.3257_{-0.0089}^{+0.0085}$ & $0.3167_{-0.0059}^{+0.0057}$\\
        ${\sigma }_{8}$ & $0.8078_{-0.0088}^{+0.0092}$ & $0.8016_{-0.0080}^{+0.0077}$ & $0.7968_{-0.0096}^{+0.0102}$ & $0.8058_{-0.0081}^{+0.0085}$\\
        ${H}_{0}~[\mathrm{km/s/Mpc}]$ & $67.52_{-0.63}^{+0.65}$ & $66.69_{-0.54}^{+0.50}$ & $66.04_{-0.88}^{+0.84}$ & $67.00_{-0.57}^{+0.60}$\\
        \hline
        ${\chi }^{2}$ & $12399.6$ & $12631.3$ & $11015.1$ & $12624.6$\\
        AIC & $12415.6$ & $12647.3$ & $11031.1$ & $12640.6$\\
        $\Delta\text{ln}\mathcal{Z}$ & -3.77 & 3.02 & 0.15 & 0.68\\
        \hline\hline

        \multicolumn{5}{l}{\textbf{LOG}} \\
        \hline
        ${w}_{0}$ & $-0.854\pm0.050$ & $-0.794\pm0.047$ & $-0.724_{-0.070}^{+0.073}$ & $-0.791_{-0.058}^{+0.055}$\\
        ${w}_{a}$ & $-0.437\pm0.145$ & $-0.543_{-0.139}^{+0.145}$ & $-0.691_{-0.174}^{+0.207}$ & $-0.587_{-0.177}^{+0.182}$\\
        \hline
        ${\Omega }_{m}$ & $0.3116_{-0.0057}^{+0.0053}$ & $0.3183\pm0.0057$ & $0.3252_{-0.0077}^{+0.0074}$ & $0.3168\pm0.0060$\\
        ${\sigma }_{8}$ & $0.8083_{-0.0085}^{+0.0088}$ & $0.8028_{-0.0083}^{+0.0077}$ & $0.7980\pm0.0091$ & $0.8059_{-0.0086}^{+0.0084}$\\
        ${H}_{0}~[\mathrm{km/s/Mpc}]$ & $67.51_{-0.58}^{+0.61}$ & $66.80\pm0.56$ & $66.11_{-0.78}^{+0.73}$ & $66.99_{-0.58}^{+0.60}$\\
        \hline
        ${\chi }^{2}$ & $12398.1$ & $12631.1$ & $11014.6$ & $12626.0$\\
        AIC & $12414.1$ & $12647.1$ & $11030.6$ & $12642.0$\\
        $\Delta\text{ln}\mathcal{Z}$ & -0.51 & 4.33 & 3.33 & 1.22 \\
        \hline\hline

        \multicolumn{5}{l}{\textbf{EXP}} \\
        \hline
        ${w}_{0}$ & $-1.062\pm0.038$ & $-1.076_{-0.040}^{+0.038}$ & $-1.073_{-0.037}^{+0.043}$ & $-1.092_{-0.042}^{+0.047}$\\
        ${w}_{a}$ & $-0.104_{-0.038}^{+0.036}$ & $-0.157_{-0.037}^{+0.043}$ & $-0.195_{-0.050}^{+0.052}$ & $-0.161\pm0.045$\\
        \hline
        ${\Omega }_{m}$ & $0.3112_{-0.0059}^{+0.0054}$ & $0.3185\pm0.0055$ & $0.3277_{-0.0089}^{+0.0082}$ & $0.3173_{-0.0062}^{+0.0058}$\\
        ${\sigma }_{8}$ & $0.8102\pm0.0088$ & $0.8069_{-0.0079}^{+0.0077}$ & $0.8012_{-0.0099}^{+0.0090}$ & $0.8107_{-0.0092}^{+0.0094}$\\
        ${H}_{0}~[\mathrm{km/s/Mpc}]$ & $67.48\pm0.60$ & $66.75\pm0.55$ & $65.85_{-0.87}^{+0.80}$ & $66.93\pm0.57$\\
        \hline
        ${\chi }^{2}$ & $12399.8$ & $12631.7$ & $11015.0$ & $12625.4$\\
        AIC & $12415.8$ & $12647.7$ & $11031.0$ & $12641.4$\\
        $\Delta\text{ln}\mathcal{Z}$ & -3.37 & 1.72 & -1.27 & -1.20\\
        \hline\hline
    \end{tabular}
\end{table}
}%

\subsubsection{$w$CDM}
In the $w$CDM model, ${w}_{0}$ is a variable constant. This model, as the smallest extension of the $\Lambda$CDM, is the only single-parameter DDE model included in our study. The parameter constraint results are shown in Figure~\ref{subfig:b}.
It is noteworthy that for CMB+BAO+DES-Y5, we find $w_0 = -0.968\pm0.018$, corresponding to a deviation of $w_0 = -1$ at 68\% C.L.. However, at 95\% C.L., $w_0 = -0.968_{-0.035}^{+0.038}$, this result includes the value $w_0 = -1$. A similar trend is observed for CMB+BAO(without LRG1 and LRG2)+DES-Y5: we get $w_0$ = $-0.972\pm0.021$ at 68\% C.L. and at 95\% C.L., the value of $w_0$ is $-0.972_{-0.044}^{+0.042}$, where $w_0 = -1$ lies outside the 1$\sigma$ interval but within the 2$\sigma$ range. The red line in Figure~\ref{rho-D5} shows the relationship between dark energy density and redshift in $w$CDM, which is described as ${\rho}_{de}(z)/{\rho}_{de}(0)=(1+z)^{3(1+w_0)}$. In the range of 1$\sigma$, the exponential term is positive and the image shows a slow monotonic increase.



Our constraints on ${w}_{0}$ show consistency with the results from \cite{DESI:2025zgx}, which also uses the DESI DR2 and three SNIa datasets, with only slight differences in the CMB data: a combination of Planck TTTEEE SimAll, Commander ($l<30$) and CamSpec ($l\ge 30$) likelihoods, including Planck and ACT DR6 lensing. We give the results as follows: in DESI+CMB+PantheonPlus, ${w}_{0}$ = $-0.995\pm0.023$, in DESI+CMB+DES-Y5, ${w}_{0}$ = $-0.971\pm0.021$, in DESI+CMB+Union3, ${w}_{0}$ = $-0.997\pm0.027$. The constraints indicate that variations for the CMB data have not significantly affected ${w}_{0}$ in the $w$CDM model. 

\par The ${H}_{0}$ value of the $w$CDM model is higher than all two-parameter DDE models, but lower than the $\Lambda$CDM model. Compared with ${H}_{0}$ = $67.36\pm0.54$ $\mathrm{km/s/Mpc}$ from Planck PR3 \cite{Planck:2018vyg} and ${H}_{0}$ = $73.04 \pm 1.04$ $\mathrm{km/s/Mpc}$ from Local \cite{Riess:2021jrx}, in CMB+BAO+PantheonPlus, ${H}_{0}$ = $67.97\pm0.50$ $\mathrm{km/s/Mpc}$, which differs by 0.83$\sigma$ from PR3 and 4.39$\sigma$ from Local; in CMB+BAO+Union3, ${H}_{0}$ = $68.04\pm0.63$ $\mathrm{km/s/Mpc}$, which differs by 0.82$\sigma$ from PR3 and 4.11$\sigma$ from Local. Compared to $\Lambda$CDM, the error bars of ${H}_{0}$ for $w$CDM are slightly larger. It is precisely this wider constraint that allows the $w$CDM to slightly alleviate the ${H}_{0}$ tension, which is relatively obvious when combined with the CMB+BAO+Union3.

Our constraints for ${\Omega }_{m}$ are $0.3045_{-0.0050}^{+0.0045}$ in CMB+BAO+PantheonPlus, $0.3101_{-0.0041}^{+0.0044}$ in CMB+BAO+DES-Y5, $0.3040\pm0.0055$ in CMB+BAO+Union3 and $0.3133_{-0.0050}^{+0.0059}$ in CMB+BAO(without LRG1 and LRG2)+DES-Y5. Compared with ${\Omega }_{m}$ = $0.3153\pm0.0073$ from Planck PR3 \cite{Planck:2018vyg}, the differences for these datasets are 1.26$\sigma$, 0.61$\sigma$, 1.24$\sigma$ and 0.21$\sigma$, respectively. \cite{Wu:2025wyk} uses the DESI DR1+DES-Y5+Cosmic Chronometers+Strong Gravitational Lensing dataset, giving ${\Omega }_{m}$ = $0.295\pm0.014$, which has a difference of 1.04$\sigma$ with ${\Omega }_{m}$ = $0.3101_{-0.0041}^{+0.0044}$ in CMB+BAO+DES-Y5. Our result has a relatively small error bar, reduced by about an order of magnitude.

\subsubsection{Chevallier-Polarski-Linder parameterization}
Chevallier-Polarski-Linder (CPL) parameterization \cite{Chevallier:2000qy, Linder:2002et} is the most basic and widely applicable two-parameter extension model, and its DE EOS is:
\begin{equation}
w(a)={w}_{0}+{w}_{a}(1-a).
\end{equation}
$X(a)$ under CPL parameterization is:
\begin{equation}
X(a)={a}^{-3-3{w}_{0}-3{w}_{a}}\text{Exp}[3{w}_{a}(a-1)]. 
\end{equation}

\par 
The CPL parameterization was proposed to address the inadequacy of the conventional first-order expansion: $w(z)={w}_{0}+{w}_{a}z$ at redshifts $z > 1$. It has a straightforward physical interpretation and provides a manageable two-dimensional phase space. Meanwhile, it has also reconstructed many scalar field EOS, and the accuracy of the resulting distance-redshift relations has been further improved.

As shown by the blue line in Figure~\ref{w(z)-D5}, at low redshifts, the CPL reduces the limitations of the earlier linear redshift behavior. The image decreases monotonically within the full redshift range and crosses the standard value $w(z)=-1$ from top to bottom. It also ensures bounded behavior at high redshifts, with an asymptotic value of ${w}_{0}+{w}_{a}$. The relationship between dark energy density and redshift in CPL can be described as ${\rho}_{de}(z)/{\rho}_{de}(0)=(1+z)^{3(1+{w}_{0}+{w}_{a})}\cdotp{e}^{-\frac{3{w}_{a}z}{1+z}}$, which is represented in Figure~\ref{rho-D5}. The curve first increases monotonically, then decreases monotonically, and crosses the reference line from top to bottom.

The ${w}_{0}$ value of CPL in all datasets can ensure that the current value of DE EOS -1 is not included in the 3$\sigma$ confidence interval. Figure~\ref{subfig:c} shows ${w}_{0}$ under the constraint of CMB+BAO+PantheonPlus as an example. Our constraint reads $w_0 = -0.841_{-0.056}^{+0.050}$ at 68\% C.L., and even at 99\% C.L. ($w_0 = -0.841_{-0.139}^{+0.137}$), the lower bound still lies above $-1$. When comparing our result of $w_0$ in CMB+BAO+PantheonPlus with $w_0 = -0.829\pm0.088$ from \cite{Shah:2024rme}, we find a discrepancy of $0.12\sigma$, showing good consistency, while the error bar is reduced by 40\%. Pay attention to another parameter ${H}_{0}$. Although the error bar of ${H}_{0}$ further increases, the value of ${H}_{0}$ decreases, which hardly alleviates ${H}_{0}$ tension.



\par 

On the other hand, our result of ${\Omega }_{m}$ under CMB+BAO+DES-Y5 (${\Omega }_{m}$ = $0.3186_{-0.0051}^{+0.0056}$ at 68\% C.L.) is 0.65$\sigma$ different from ${\Omega }_{m}$ = $0.329_{-0.015}^{+0.018}$ \cite{Wu:2025wyk}. We provide stricter constraints, which reduce the error bar by 68\%, and further lower the value of $\Omega_m$. However, \cite{DESI:2025zgx} provides tight constraints on ${\Omega }_{m}$. Here are the results: in DESI+CMB+PantheonPlus, ${\Omega }_{m}$ = $0.3114\pm0.0057$, in DESI+CMB+DES-Y5, ${\Omega }_{m}$ = $0.3191\pm0.0056$, and in DESI+CMB+Union3, ${\Omega }_{m}$ = $0.3275\pm0.0086$. This is in good agreement with our corresponding results, and the difference between them can be ignored. 

The high sensitivity of CPL to observational data allows a better study of the time-varying EOS component during the CMB decoupling era. However, the model diverges as the redshift $z$ approaches -1. This issue has motivated the development of other parameterization forms.

\subsubsection{Jassal-Bagla-Padmanabhan parameterization}
We consider the Jassal-Bagla-Padmanabhan (JBP) \cite{Jassal:2005qc} form for the DE EOS:
\begin{equation}
w(a)={w}_{0}+{w}_{a}a(1-a).
\end{equation}
Its $X(a)$ is :
\begin{equation}
X(a)={a}^{-3-3{w}_{0}}\text{Exp}\left[\frac{3}{2}{w}_{a}{(a-1)}^{2}\right]. 
\end{equation}

\par 
The JBP parameterization allows for greater flexibility in dark energy evolution. This improves its fit with the observational data. The orange line in Figure~\ref{w(z)-D5} shows the variation characteristics of $w(z)$ in the JBP model. In both low and high redshifts, the limiting value of the JBP is ${w}_{0}$. Its minimum occurs at $z = 1$, with a value of $w(z)=w_{0}+\frac{1}{4}w_{a}$. Thus, it can describe a dark energy component with the same EOS at present and remote past. In low redshifts, the graph of $w(z)$ changes rapidly, facilitating the detection of fast evolution at small redshifts. At $z\gg1$, observations are less sensitive to changes in $w(z)$. In this time, returning to the present value is not crucial. In the following, we also give the expression of ${\rho}_{de}(z)/{\rho}_{de}(0)$, and the result is $(1+z)^{3(1+{w}_{0})}\cdotp{e}^{\frac{3}{2}w_a(\frac{1}{1+z}-1)^2}$. The image is shown in Figure~\ref{rho-D5}. The curve crosses the standard line twice and has three intersections.

 From Figure~\ref{subfig:d}, we find that ${\Omega }_{m}$ has an obvious inverse correlation with both ${H}_{0}$ and ${\sigma }_{8}$, while ${H}_{0}$ and ${\sigma }_{8}$ show a positive correlation. Meanwhile, $w_0$ and $w_a$ also exhibit an inverse proportional relationship. The value of $w_0$ from CMB+BAO+PantheonPlus is $-0.813_{-0.148}^{+0.158}$ at 95\% C.L., and $-0.813\pm0.196$ at 99\% C.L.. This indicates that $w_0 = -1$ lies outside the 2$\sigma$ confidence interval but within the 3$\sigma$ interval. For all other datasets, $w_0$ can deviate from $-1$ even at 99\% C.L..

Among all DDE parameterizations discussed in this work, JBP shows the largest deviation of $w_0$ and $w_a$ from their standard values ($w_0 = -1$, $w_a = 0$). Taking the constraints from CMB+BAO+Union3 as an example, we obtain $w_0 = -0.542_{-0.129}^{+0.115}$ and $w_a = -2.421_{-0.613}^{+0.685}$ at 68\% C.L.. Both parameters can deviate from their respective standard values at 3$\sigma$ level. \cite{Chaudhary:2025vzy} also exhibits constraints of the DE EOS parameters when combined with Planck 2018+DESI DR2+Union3. Specifically, their ${w}_{0}$ = $-0.621_{-0.093}^{+0.130}$ differs from ours by 0.43$\sigma$, and their ${w}_{a}$ = $-1.860_{-0.730}^{+0.490}$ is 0.56$\sigma$ higher than our result. The CMB data that we used are different, which may lead to this discrepancy. Compared to the $\Lambda$CDM model, the JBP model yields lower ${H}_{0}$ values when constrained by all datasets, which undoubtedly exacerbates ${H}_{0}$ tension. 


\par 

\par 


Our results, based on DESI DR2 and CMB PR4, provide relatively tighter ${\sigma }_{8}$ constraints and slightly alleviate ${\sigma }_{8}$ tension. Specifically, we obtain ${\sigma }_{8}$ = $0.7977_{-0.0089}^{+0.0083}$ at 68\% C.L. in CMB+BAO+DES-Y5. Our result is 1.24$\sigma$ lower than ${\sigma }_{8}$ = $0.819\pm0.015$ \cite{ACT:2023kun}, which is provided by the combination of ACT DR6 lensing and BAO. Compared with ${\sigma }_{8}$ = $0.8111\pm0.0060$ from Planck PR3 TTTEEE+lowE+lensing \cite{Planck:2018vyg}, the difference is 1.31$\sigma$. As for DES-Y3, the value of ${\sigma }_{8}$ is $0.783_{-0.092}^{+0.073}$ \cite{DES:2021bvc, DES:2021vln} and the difference is 0.20$\sigma$. Although the difference is small, the main reason is still the relatively broad error bar. In \cite{Giare:2024gpk}, the result given is ${\sigma }_{8}$ = $0.8083\pm0.0086$ under Planck 2018+DESI DR1+DES-Y5. In comparison, our result is 0.89$\sigma$ lower when the error bar is similar, slightly reducing the ${\sigma }_{8}$ tension.

 
JBP adopts this point of view: if dark energy is not a key dynamical factor at $z\gtrsim1$, we should focus on $z\lesssim1$. Compared to CPL, JBP behaves differently at low redshifts. This is because its form combines linear and quadratic terms of the scale factor. At present, the quadratic term $-{w}_{a}\cdotp a^2$ becomes comparable to the linear term ${w}_{a}\cdotp a$. Similarly, this parameterization also diverges in the future as $z$ approaches -1, like CPL.

\subsubsection{Feng–Shen–Li–Li parameterization}
The next model we study is the Feng–Shen–Li–Li (FSLL) parameterization \cite{Feng:2012gf}. It has two cases:
\begin{equation}
\text{FSLL I}:w(a)={w}_{0}-{w}_{a}\frac{a(1-a)}{{a}^{2}+{(a-1)}^{2}},
\end{equation}
\begin{equation}
\text{FSLL II}:w(a)={w}_{0}-{w}_{a}\frac{{(a-1)}^{2}}{{a}^{2}+{(a-1)}^{2}}.
\end{equation}
Corresponding $X(a)$ can be described as:
\begin{equation}
\text{FSLL I}:X(a)={a}^{-3-3{w}_{0}}\text{Exp}\left[\frac{3}{8}{w}_{a}(\pi +4\arctan (1-2a)+2\ln_{}{(2{a}^{2}-2a+1)})\right], 
\end{equation}
\begin{equation}
\text{FSLL II}:X(a)={a}^{-3-3{w}_{0}-3{w}_{a}}\text{Exp}\left[-\frac{3}{8}{w}_{a}(\pi +4\arctan (1-2a)-2\ln_{}{(2{a}^{2}-2a+1)})\right].
\end{equation}

\par 
As mentioned above, the CPL exhibits a divergence issue when $z$ approaches -1, which undoubtedly represents a non-physical feature. This hinders the CPL from truly encompassing scalar field models and other theoretical models. The FSLL models overcome this issue, extending the parameterization of dark energy to redshifts $z$ approaching -1, and ensuring that they do not diverge in the whole range of redshifts $z\in[-1,\infty)$. 

In Figure~\ref{w(z)-D5}, the brown line represents the FSLL I model, while the pink line corresponds to the FSLL II model. Specifically, when $z=-1$, FSLL I has $w(z)=w_{0}-\frac{1}{2}w_{a}$, while FSLL II has $w(z)=w_{0}+\frac{1}{2}w_{a}$. FSLL I is similar to JBP and is sensitive to changes in low redshifts. Whether $z=0$ or $z\gg1$, the asymptotic value is $w_{0}$. When $z=1$, $w(z)$ takes the minimum value: $w_{0}+\frac{1}{2}w_{a}$. It is not difficult to find that for $z\ll1$, FSLL I will be simplified to the linear form: $w(z)\approx{w}_{0}+{w}_{a}\cdotp z$ as CPL. FSLL II is closer to CPL, and $w(z)$ is taken as ${w}_{0}$ at $z=0$, while the asymptotic value at high redshifts is ${w}_{0}+{w}_{a}$. Similarly, for $z\ll1$, FSLL II reduces to the quadratic form: $w(z)\approx{w}_{0}+{w}_{a}\cdotp z^2$.

The expressions ${\rho}_{de}(z)/{\rho}_{de}(0)$ of FSLL I and FSLL II are relatively complex. The results are as follows:
\begin{equation}
\text{FSLL I}: {\rho}_{de}(z)/{\rho}_{de}(0) = (1+z)^{3(1+{w}_{0})}\cdotp{e}^{\frac{3}{8}{w}_{a}(\pi+4\arctan(1-\frac{2}{1+z})+2\ln(2(\frac{1}{1+z})^2-2(\frac{1}{1+z})+1))},
\end{equation}
\begin{equation}
\text{FSLL II}: {\rho}_{de}(z)/{\rho}_{de}(0) = (1+z)^{3(1+{w}_{0}+{w}_{a})}\cdotp{e}^{-\frac{3}{8}{w}_{a}(\pi+4\arctan(1-\frac{2}{1+z})-2\ln(2(\frac{1}{1+z})^2-2(\frac{1}{1+z})+1))}.
\end{equation}
The specific images of the two models are shown in Figure~\ref{rho-D5}. Our results show that the curve of FSLL I is similar to that of JBP, but JBP is steeper at high redshifts. The curve of FSLL II resembles that of CPL: at low redshifts, FSLL II changes slightly more rapidly, while at high redshifts, it tends to converge with CPL.

The DE EOS parameters of FSLL I and FSLL II deviate from their respective standard values at 99\% C.L. across all datasets. We present constraints on the DE EOS parameters for both models under the CMB+BAO+PantheonPlus, as the deviation from the standard values is smaller than for other datasets. To be specific, for FSLL I, we find $w_0 = -0.844_{-0.143}^{+0.153}$ and $w_a = -0.543_{-0.505}^{+0.521}$ at 99\% C.L.; for FSLL II, the corresponding constraints are $w_0 = -0.898_{-0.091}^{+0.088}$ and $w_a = -0.565_{-0.510}^{+0.410}$ at 99\% C.L.. Figures~\ref{subfig:e} and~\ref{subfig:f} show the triangular plots of FSLL I and FSLL II respectively. Combined with the data, we can get that the $w_0$ parameters of the two models are similar under CMB+BAO+PantheonPlus, but FSLL II gives a smaller error bar.

We have strict constraints on the DE EOS parameters. In comparison, \cite{Staicova:2022zuh} analyzes two forms of the FSLL model, using a combination of different BAO with $0.11\le z\le 2.40$, CMB distant prior, the SNIa Pantheon, and the GRB dataset consisting of 162 measurements with $0.03351\le z\le 9.3$. The DE EOS parameter constraints of the two models in this paper are very similar. We only take FSLL II as an example. When only using CMB+BAO (hereinafter CB), FSLL II has ${w}_{0}$ = $-0.91 \pm 0.09$ and ${w}_{a}$ = $0.26 \pm 0.2$. When adding SNIa and GRB (hereinafter CBSG), the result of ${w}_{0}$ decreases, ${w}_{0}$ = $-1.04 \pm 0.03$ for FSLL II; while the value of ${w}_{a}$ increases, ${w}_{a}$ = $0.35 \pm 0.13$ for FSLL II. 
In our analysis of CMB+BAO+PantheonPlus, FSLL II has ${w}_{0}$ = $-0.898_{-0.036}^{+0.034}$ (68\% C.L.) with a 0.12$\sigma$ discrepancy to CB, 3.03$\sigma$ to CBSG, and ${w}_{a}$ = $-0.565_{-0.177}^{+0.193}$ (68\% C.L.) with a 2.97$\sigma$ discrepancy to CB, 3.93$\sigma$ to CBSG. Obviously, CB agrees better with our results than CBSG, which may be due to the expansion of PantheonPlus and the influence of GRB.

\subsubsection{Barboza-Alcaniz parameterization}
The Barboza-Alcaniz (BA) parameterization \cite{Barboza:2008rh} is characterized by the following form:
\begin{equation}
w(a)={w}_{0}+{w}_{a}\frac{1-a}{{a}^{2}+{(1-a)}^{2}}. 
\end{equation}
$X(a)$ can be written as:
\begin{equation}
X(a)={a}^{-3-3{w}_{0}-3{w}_{a}}{(2{a}^{2}-2a+1)}^{\frac{3}{2}{w}_{a}}. 
\end{equation}

The BA parameterization allows for a more complex evolution in the DE EOS. It can be regarded as a further extension of the CPL. When DE EOS is expressed as $w(z)={w}_{0}+\frac{z{(1+z)}^{n-1}}{1+z^n}{w}_{a}$, setting $n=1$ will degenerate into the CPL, and setting $n=2$ will get the BA. This form allows for deviations from the CPL scenario.

As shown in the purple line of Figure~\ref{w(z)-D5}, at low redshifts, this parameterization behaves linearly and the limit value ${w}_{0}$ is taken at $z=0$. During high redshifts, $w(z)$ approaches the asymptotic value ${w}_{0}+{w}_{a}$. The minimum occurs at $z=1+\sqrt{2}$, and the value is $w(z)={w}_{0}+\frac{4+3\sqrt{2}}{4+2\sqrt{2}}{w}_{a}\approx{w}_{0}+1.207{w}_{a}$. Meanwhile, when $z$ approaches -1, $w(z)$ converges to ${w}_{0}$, which ensures its boundedness in the future. For BA parameterization, the relationship between dark energy density and redshift can be expressed as ${\rho}_{de}(z)/{\rho}_{de}(0)=(1+z)^{3(1+{w}_{0}+{w}_{a})}\cdotp(1-\frac{2z}{(1+z)^2})^{\frac{3}{2}w_a}$, and the image is shown in Figure~\ref{rho-D5}. The curve of BA is similar to that of CPL, but slightly different. The BA curve is gentler, and its approach to the asymptotic value of 0 is slower at high redshifts. The reason for this may be that the expression for $\rho_{\rm de}(z)/\rho_{\rm de}(0)$ in BA is a power function, whereas that in CPL involves an exponential function.

The ${w}_{0}$ and ${w}_{a}$ of BA can also deviate from their standard values at 3$\sigma$ level in all datasets. Figure~\ref{subfig:g} shows an inverse relationship between ${w}_{0}$ and ${w}_{a}$. It is not difficult to find that the DE EOS parameters under CMB+BAO+Union3 deviate from the standard values to a large extent. Here we present the specific results: $w_0 = -0.730_{-0.184}^{+0.178}$ and $w_a = -0.483_{-0.321}^{+0.345}$ at 99\% C.L..



When comparing the DE EOS parameters of BA with \cite{Kumar:2024soe}, we find an interesting phenomenon. Our constraints on DE EOS from CMB+BAO+DES-Y5 are
$w_0 = -0.791\pm0.044$ and $w_a = -0.396_{-0.090}^{+0.103}$ at 68\% C.L.; for CMB+BAO+Union3, we obtain $w_0 = -0.730_{-0.073}^{+0.077}$ and $w_a = -0.483\pm0.127$ at 68\% C.L. \cite{Kumar:2024soe} uses BAO measurements from SDSS-IV eBOSS and combines them with three SNIa, respectively. Compared with our corresponding results, in DES-Y5+BAO, they obtain $w_0 = -0.962_{-0.036}^{+0.031}$ with a $3.18\sigma$ difference and $w_a = 0.06_{-0.34}^{+0.43}$ with a $1.28\sigma$ difference; in Union3+BAO, they show $w_0 = -0.768\pm0.080$ with a difference of $0.35\sigma$ and $w_a = -0.43_{-0.39}^{+0.49}$ with a difference of $0.13\sigma$. The latter shows better consistency, and the difference is significantly smaller than the former. This may be because Union3 has a stronger constraint effect on DE EOS of the BA, thus weakening the influence of other data. From the perspective of constraint capability, we also provide a smaller error bar for ${w}_{a}$.

Like the FSLL, the DE EOS of BA performs well over the entire cosmic expansion history. It is a completely smooth function without singularities. It should be mentioned that the BA model is considered to be better than the CPL or JBP in reducing low-redshift errors \cite{Colgain:2021pmf}.

\subsubsection{Logarithmic parameterization}
Going a step further, the logarithmic (LOG) parameterization \cite{Efstathiou:1999tm} is:
\begin{equation}
w(a)={w}_{0}-{w}_{a}\ln_{}{a}. 
\end{equation}
We can also obtain:
\begin{equation}
X(a)={a}^{-3-3{w}_{0}+\frac{3}{2}{w}_{a}\ln_{}{a}}. 
\end{equation}

\par 
As depicted by the green line in Figure~\ref{w(z)-D5}, the LOG model is a relatively special kind of parameterization and its DE EOS $w(z)$ is decreasing monotonically. In the period of low redshifts, when $z=0$, $w(z)$ takes ${w}_{0}$. As $z\rightarrow\infty$, $w(z)$ tends to infinity. This makes LOG the only parameterization discussed in our paper that exhibits unbounded behavior in the early universe. Therefore, the LOG is more suitable for describing dark energy at low redshifts. It is particularly effective in capturing the behaviors of various potential scalar field models at $z\lesssim4$ \cite{Efstathiou:1999tm}. 

Moreover, ${\rho}_{de}(z)/{\rho}_{de}(0)$ of the LOG is expressed as $(1+z)^{3(1+{w}_{0})-\frac{3}{2}w_a\ln(\frac{1}{1+z})}$, which is shown in Figure~\ref{rho-D5}. The curve of LOG shows the same characteristics as that of CPL, but there are subtle differences. In contrast to BA, LOG evolves slightly more rapidly than CPL at low redshifts, while they tend to be the same in high redshifts.

Figure~\ref{subfig:h} shows the parameter constraints of the LOG parameterization. Similarly, the DE EOS parameters of LOG in all datasets deviate from the standard value at the 3$\sigma$ level. Combined with the data, we can see that ${w}_{0}$ of LOG is similar to that of BA, but ${w}_{a}$ is slightly different. Compared to BA, ${w}_{a}$ of LOG deviates from the standard value of 0 to a larger extent, which is relatively obvious under the constraint of CMB+BAO+Union3. Our constraints yield $w_0 = -0.724_{-0.174}^{+0.196}$ and $w_a = -0.691_{-0.535}^{+0.470}$ at 99\% C.L. in CMB+BAO+Union3.
Our results are supported to some extent by \cite{Chaudhary:2025vzy}.
For the three datasets considered, the differences in the DE EOS parameters are all within $1\sigma$. Taking CMB+BAO+DES-Y5 as an example, we obtain $w_0 = -0.794\pm0.047$ and $w_a = -0.543_{-0.139}^{+0.145}$ at 68\% C.L.. Meanwhile, \cite{Chaudhary:2025vzy} also presents constraints on $w_0$ and $w_a$ for the LOG model under the Planck 2018+DESI DR2+DES-Y5 dataset. For direct comparison, their results are $w_0 = -0.822\pm0.055$ (with a difference of $0.39\sigma$) and $w_a = -0.430_{-0.160}^{+0.190}$ (with a difference of $0.52\sigma$) relative to ours. In general, these results are in good agreement with each other. It should be noted that the error bar of $w_0$ is quite similar between the two studies, while our constraint on $w_a$ produces a slightly smaller error bar, which is reduced by 19\% under the CMB+BAO+DES-Y5.

However, this does not mean that the LOG cannot describe the behavior at high redshifts. In fact, when considering the current data limitations and the logarithmic nature of the DE EOS, the LOG can also be safely extended to high redshifts. Since the data used in this work include late-time observations, the DE EOS parameter $w_a$ is tightly constrained to the region $w_a < 0$. Consequently, in the early universe, compared to other major components of cosmic energy (such as radiation and matter), the contribution of DE is still negligible to a large extent and will not affect early cosmology \cite{Giare:2024gpk, Yang:2021flj}. 

\subsubsection{Exponential parameterization}
The exponential (EXP) parameterization \cite{Dimakis:2016mip, Pan:2019brc} has the following form:
\begin{equation}
w(a)={w}_{0}+{w}_{a}\left[\text{Exp}(1-a)-1\right]. 
\end{equation}
We can get its $X(a)$ as:
\begin{equation}
X(a)={a}^{-3-3{w}_{0}+3{w}_{a}}\text{Exp}\left[3\text{e}{w}_{a}(Ei(-1)-Ei(-a))\right]. 
\end{equation}
Here $Ei(x)=-\int_{-x}^{\infty}\frac{{e}^{-x}}{x}dx$ (for $x<0$) is the exponential integral function.\\

\par 
Performing a Taylor expansion on EXP yields the following form: $w(a)={w}_{0}+{w}_{a}[(1-a)+\frac{(1-a)^2}{2!}+\frac{(1-a)^3}{3!}+\frac{(1-a)^4}{4!}+\cdots]$. It can be seen that the EXP is a further extension of the CPL. The CPL only includes the linear term in $(1-a)$, while the EXP contains all higher-order terms, allowing for small corrections beyond linear behavior.

The DE EOS $w(z)$ of EXP parameterization has a good bounded behavior throughout the history of the universe. As shown by the gray line in Figure~\ref{w(z)-D5}, at low redshifts, when $z=0$, the value of $w(z)$ is ${w}_{0}$. In the high-redshift limit ($z\rightarrow\infty$), $w(z)$ approaches its asymptotic value ${w}_{0}+(e-1){w}_{a}$, which ensures that it does not diverge in the early universe. Furthermore, $w(z)$ is also bounded when $z$ approaches -1 in the future, with a limiting value of ${w}_{0}-{w}_{a}$. 

The relationship between dark energy density and redshift of EXP is also special, as shown in Figure~\ref{rho-D5}. We give the specific expression as follows: ${\rho}_{de}(z)/{\rho}_{de}(0)=(1+z)^{3(1+{w}_{0}-w_a)}\cdotp{e}^{3ew_a[Ei(-1)-Ei(-\frac{1}{1+z})]}$. Unlike any of the models discussed above, the curve of EXP decreases monotonously in the range of $z>0$, and does not cross the standard line ${\rho}_{de}(z)/{\rho}_{de}(0)=1$. It can be seen that, at low redshifts, the curve evolves more slowly than in the CPL and also approaches 0 at high redshifts.

We can find in Figure~\ref{subfig:i} that ${w}_{0}$ and ${w}_{a}$ of EXP do not show an inverse relationship like other parameterizations but rather tend to be proportional. Moreover, ${w}_{0}$ is also relatively smaller than other models. The value of ${w}_{0}$ under CMB+BAO(without LRG1 and LRG2)+DES-Y5 is $-1.092_{-0.093}^{+0.084}$ at 95\% C.L., and the upper limit is less than -1. ${w}_{0}$ in other datasets can also be less than -1 at 68\% C.L.. The ${w}_{a}$ of EXP also has unique features compared to other models: its error bar is relatively smaller. However, ${w}_{a}$ can deviate from 0 at 99\% C.L. in all datasets, which is consistent with other models.




The constraints of ${w}_{0}$ from EXP have little difference in the four datasets, and the deviation of ${w}_{a}$ from the standard value is relatively large under CMB+BAO+Union3. Taking the results under CMB+BAO+Union3 as an example, we find ${w}_{0}=-1.073_{-0.037}^{+0.043}$ and ${w}_{a}=-0.195_{-0.050}^{+0.052}$ at 68\% C.L.. In comparison, \cite{Arora:2025msq} also gives the results of ${w}_{0}$ and ${w}_{a}$ for the EXP model under the joint constraints of CMB and DESI DR2. The results are ${w}_{0}$ = $-0.51_{-0.22}^{+0.17}$ and ${w}_{a}$ = $-1.19_{-0.57}^{+0.55}$. Compared to our results, ${w}_{0}$ has a discrepancy of 2.51$\sigma$ and ${w}_{a}$ has a discrepancy of 1.80$\sigma$. Thus, the difference may be mainly due to Union3. For EXP, the use of SNIa data makes ${w}_{0}$ decrease and ${w}_{a}$ increase. Meanwhile, the addition of SNIa data has also improved the error bar.

According to previous experience, a purely linear parameterization is a relatively rigid form. It lacks flexibility and may not accurately describe the dynamics of dark energy throughout the cosmic expansion history. The EXP addresses this limitation to some extent. It can explore whether neighboring higher-order terms are significant without increasing the dimensionality of the parameter space.
\\
\par The differences mentioned above are mainly due to updates and expansions of the datasets, which enhance the constraints on the models and alter the posterior distribution of the parameter space in various ways. Firstly, the update to CMB NPIPE (PR4) CamSpec plays a key role. NPIPE is a new independent pipeline that offers substantial improvements in detector calibration and system correction compared to previous versions. Its low noise allows for tighter parameter constraints, and due to improved polarization, most $\Lambda$CDM parameters in TTTEEE have also improved by approximately 10\% \cite{Rosenberg:2022sdy}. Secondly, updates to DESI DR2 have also played a crucial role. Compared to DESI DR1, the effective data volume has more than doubled and the statistical accuracy has also been improved. With more than 30 million redshifts for galaxies and quasars, and Ly$\alpha$ forest spectra for over 820,000 quasars, DESI DR2 represents the largest spectroscopic galaxy sample so far and provides the most precise BAO measurements at any redshift \cite{DESI:2025zgx}.
\begin{figure*}[htbp]
    \centering
    \begin{tabular}{ccc}
        \begin{subfigure}[b]{0.3\textwidth}
            \centering
            \includegraphics[width=\linewidth]{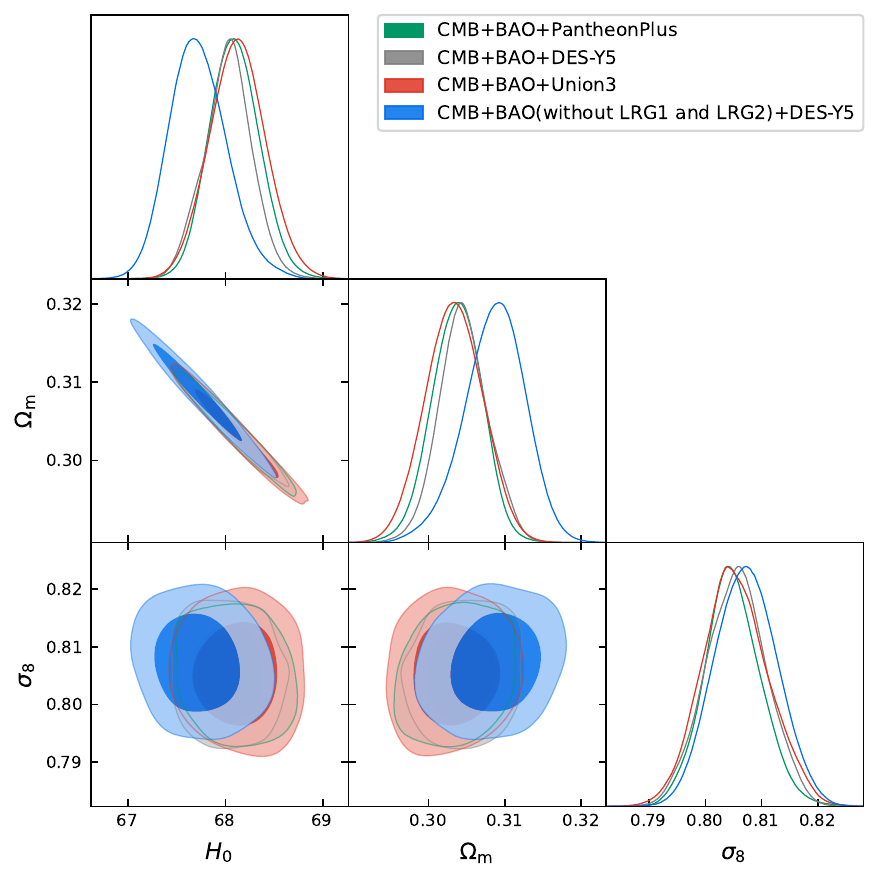}  
            \caption{$\Lambda$CDM Model}
            \label{subfig:a}
        \end{subfigure}
        &
        \begin{subfigure}[b]{0.3\textwidth}
            \centering
            \includegraphics[width=\linewidth]{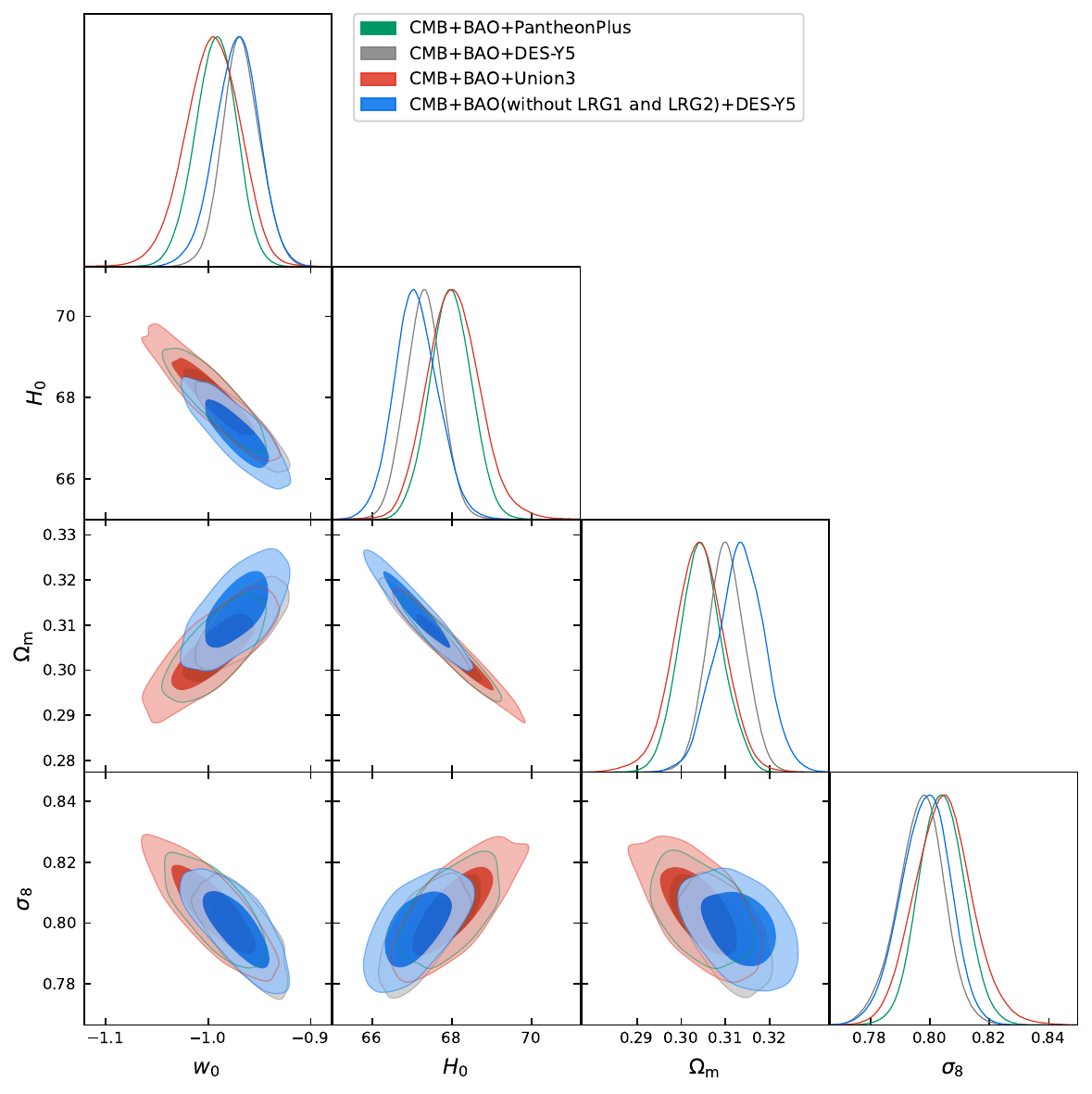}  
            \caption{$w$CDM Parameterization}
            \label{subfig:b}
        \end{subfigure}
        &
        \begin{subfigure}[b]{0.3\textwidth}
            \centering
            \includegraphics[width=\linewidth]{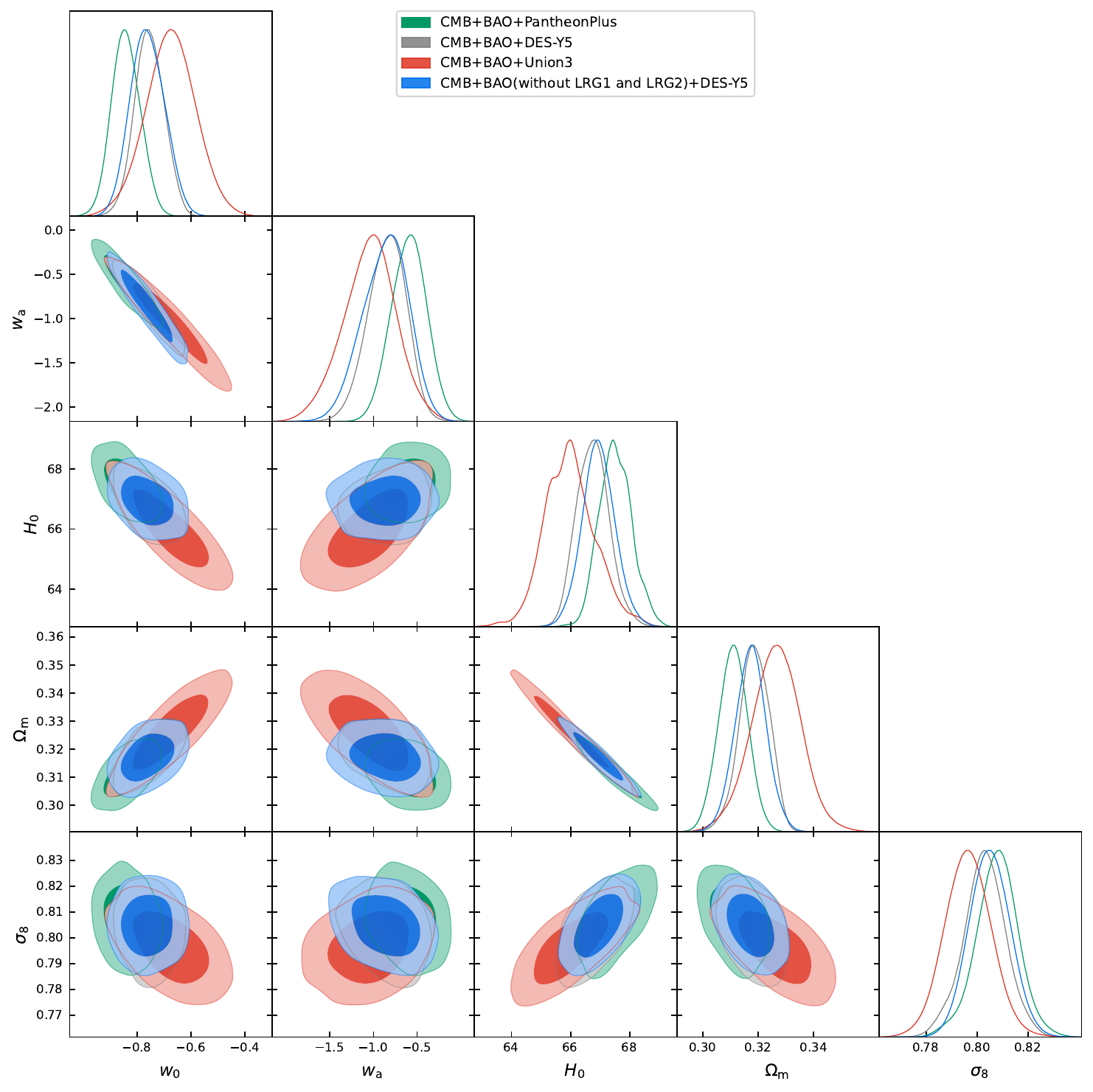}  
            \caption{CPL Parameterization}
            \label{subfig:c}
        \end{subfigure}
        \\[10pt]  
        
        \begin{subfigure}[b]{0.3\textwidth}
            \centering
            \includegraphics[width=\linewidth]{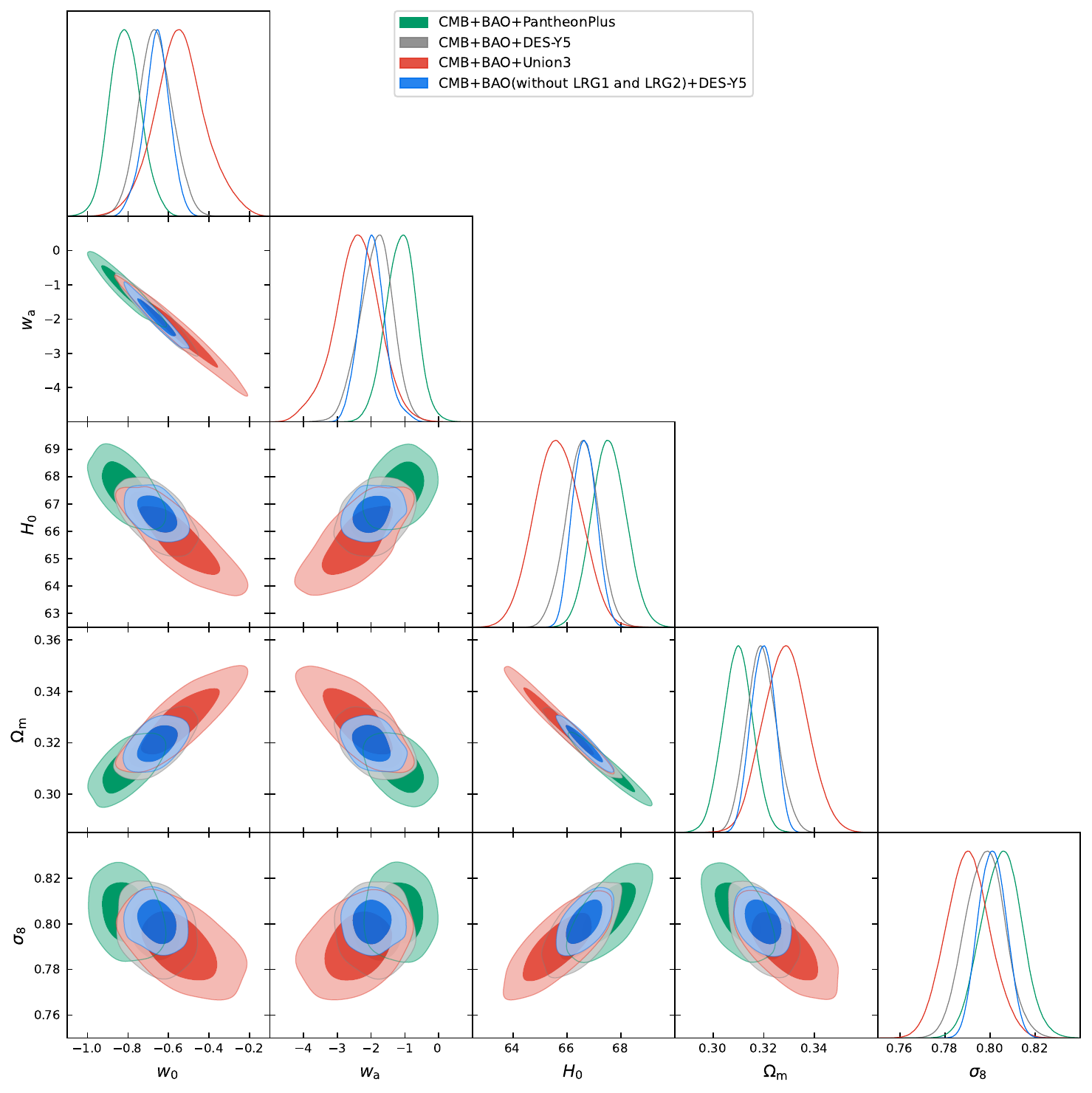}  
            \caption{JBP Parameterization}
            \label{subfig:d}
        \end{subfigure}
        &
        \begin{subfigure}[b]{0.3\textwidth}
            \centering
            \includegraphics[width=\linewidth]{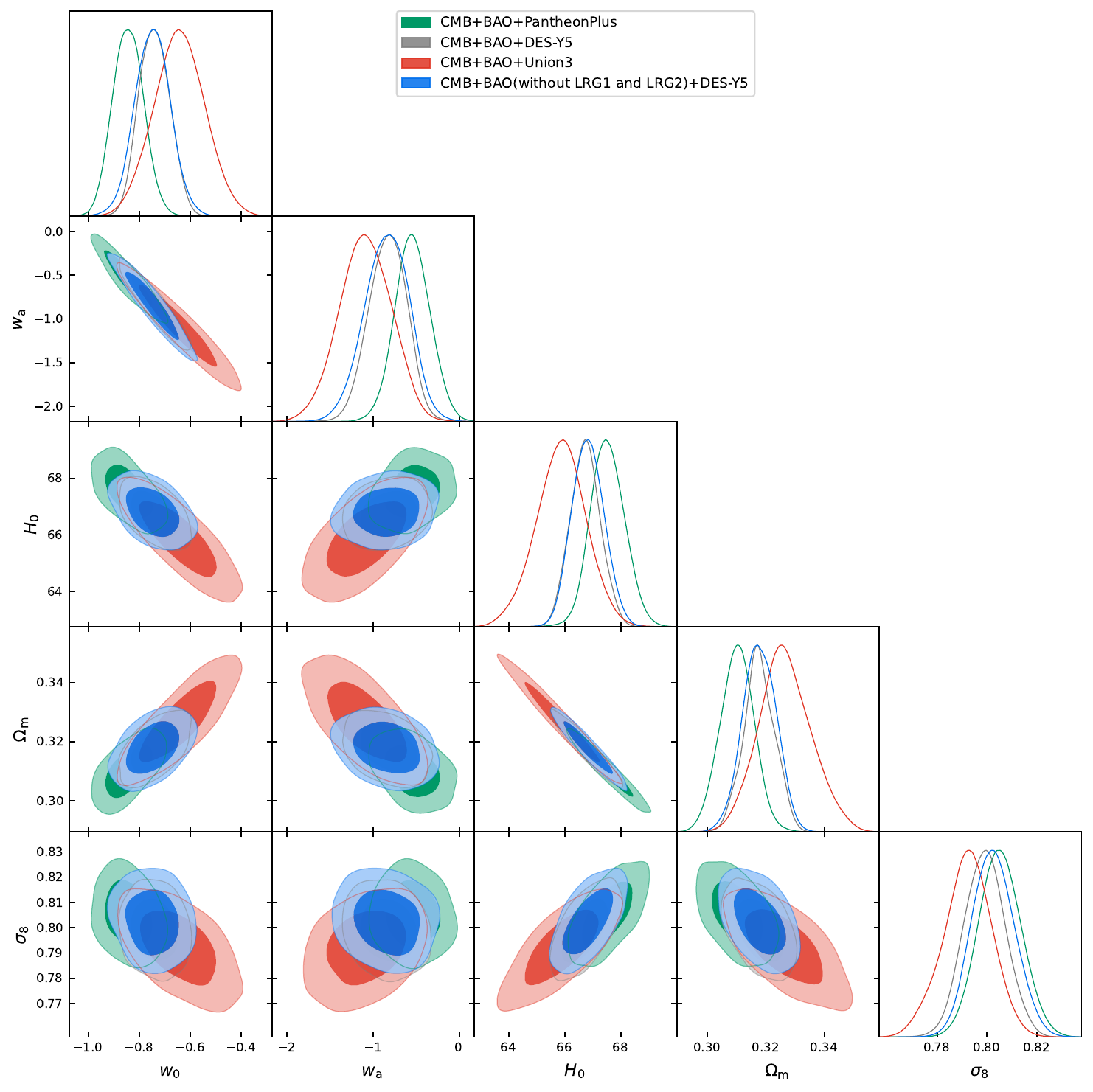}  
            \caption{FSLL I Parameterization}
            \label{subfig:e}
        \end{subfigure}
        &
        \begin{subfigure}[b]{0.3\textwidth}
            \centering
            \includegraphics[width=\linewidth]{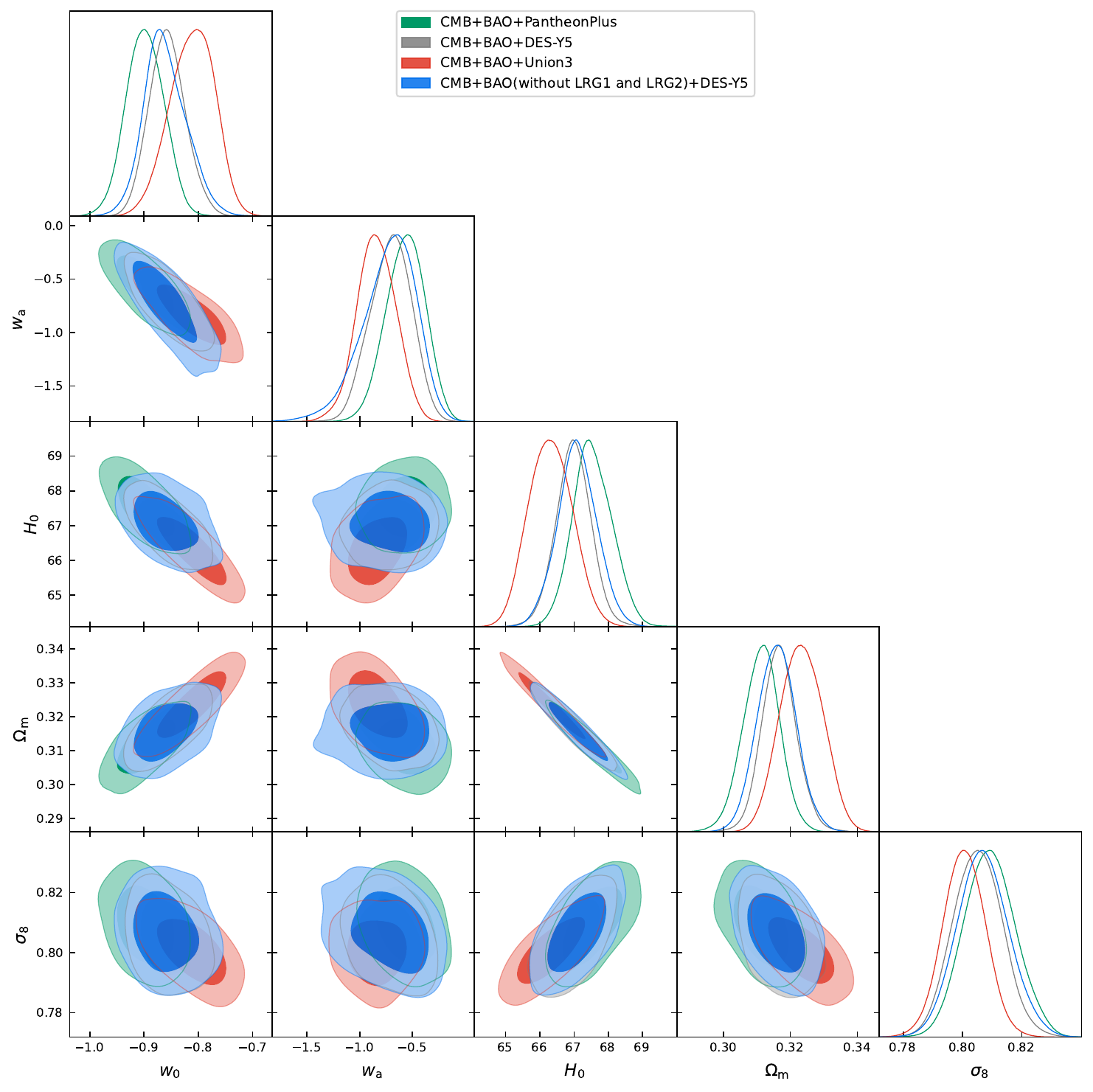}  
            \caption{FSLL II Parameterization}
            \label{subfig:f}
        \end{subfigure}
        \\[10pt]  
        
        \begin{subfigure}[b]{0.3\textwidth}
            \centering
            \includegraphics[width=\linewidth]{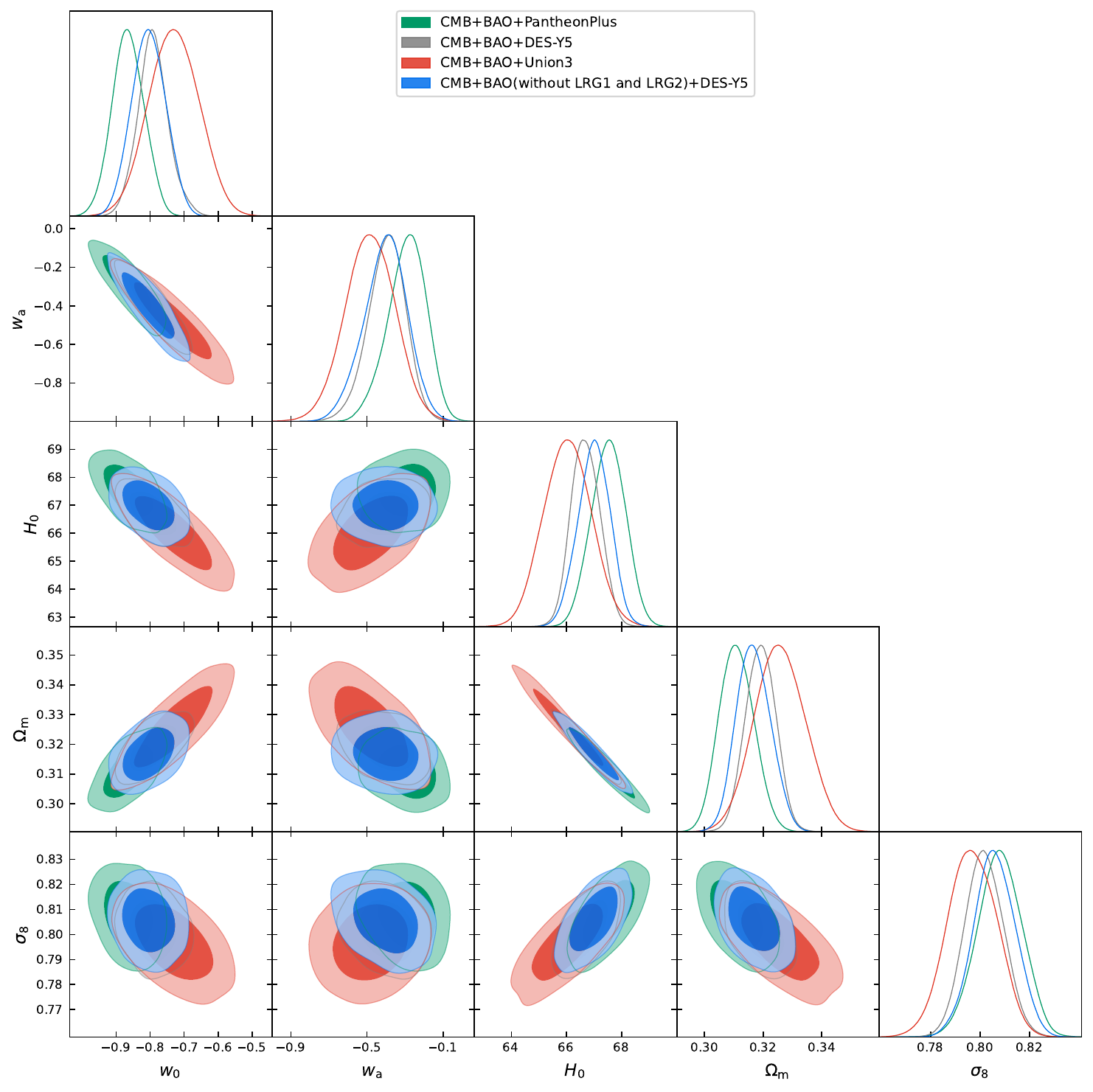}  
            \caption{BA Parameterization}
            \label{subfig:g}
        \end{subfigure}
        &
        \begin{subfigure}[b]{0.3\textwidth}
            \centering
            \includegraphics[width=\linewidth]{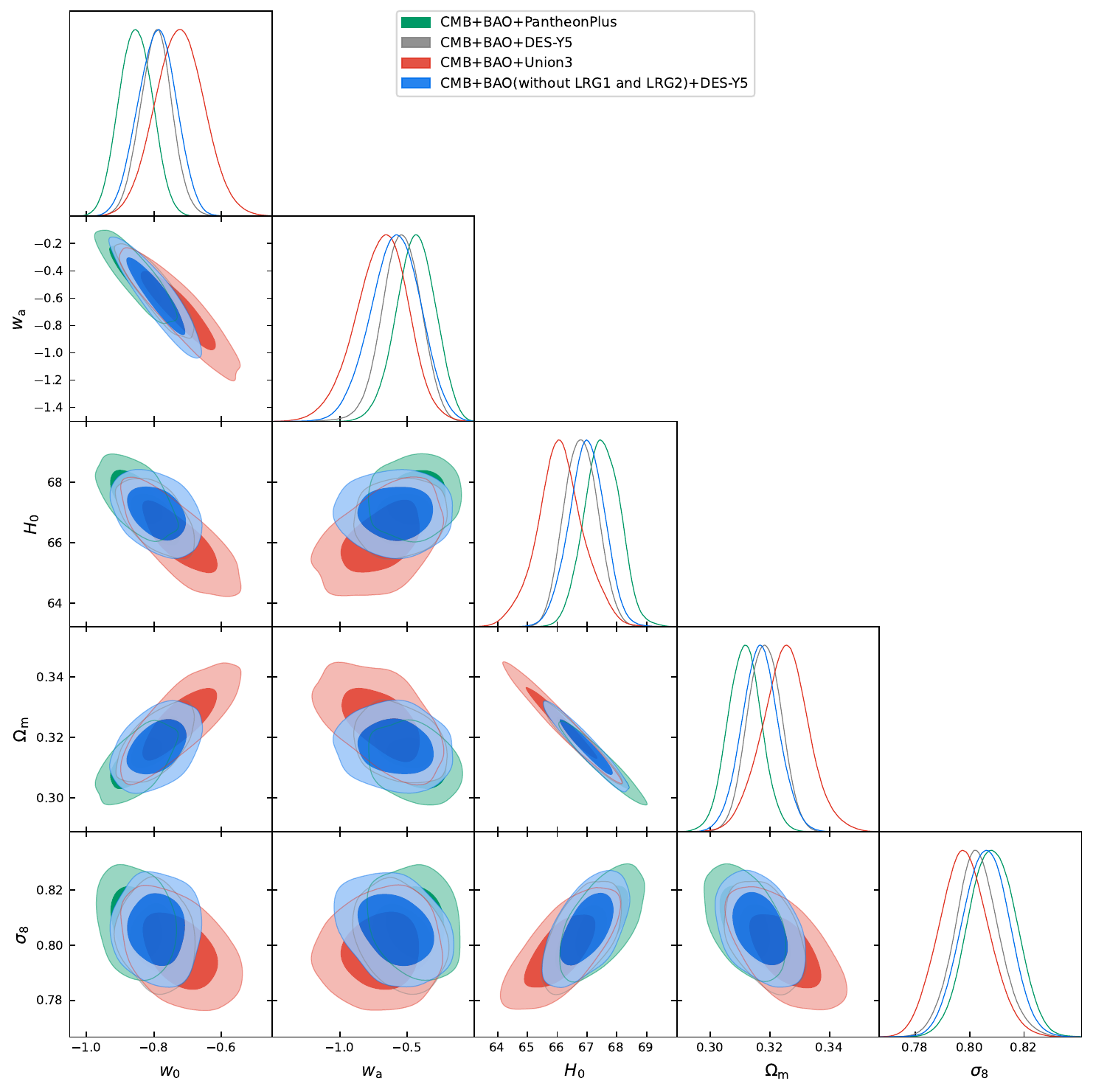}  
            \caption{LOG Parameterization}
            \label{subfig:h}
        \end{subfigure}
        &
        \begin{subfigure}[b]{0.3\textwidth}
            \centering
            \includegraphics[width=\linewidth]{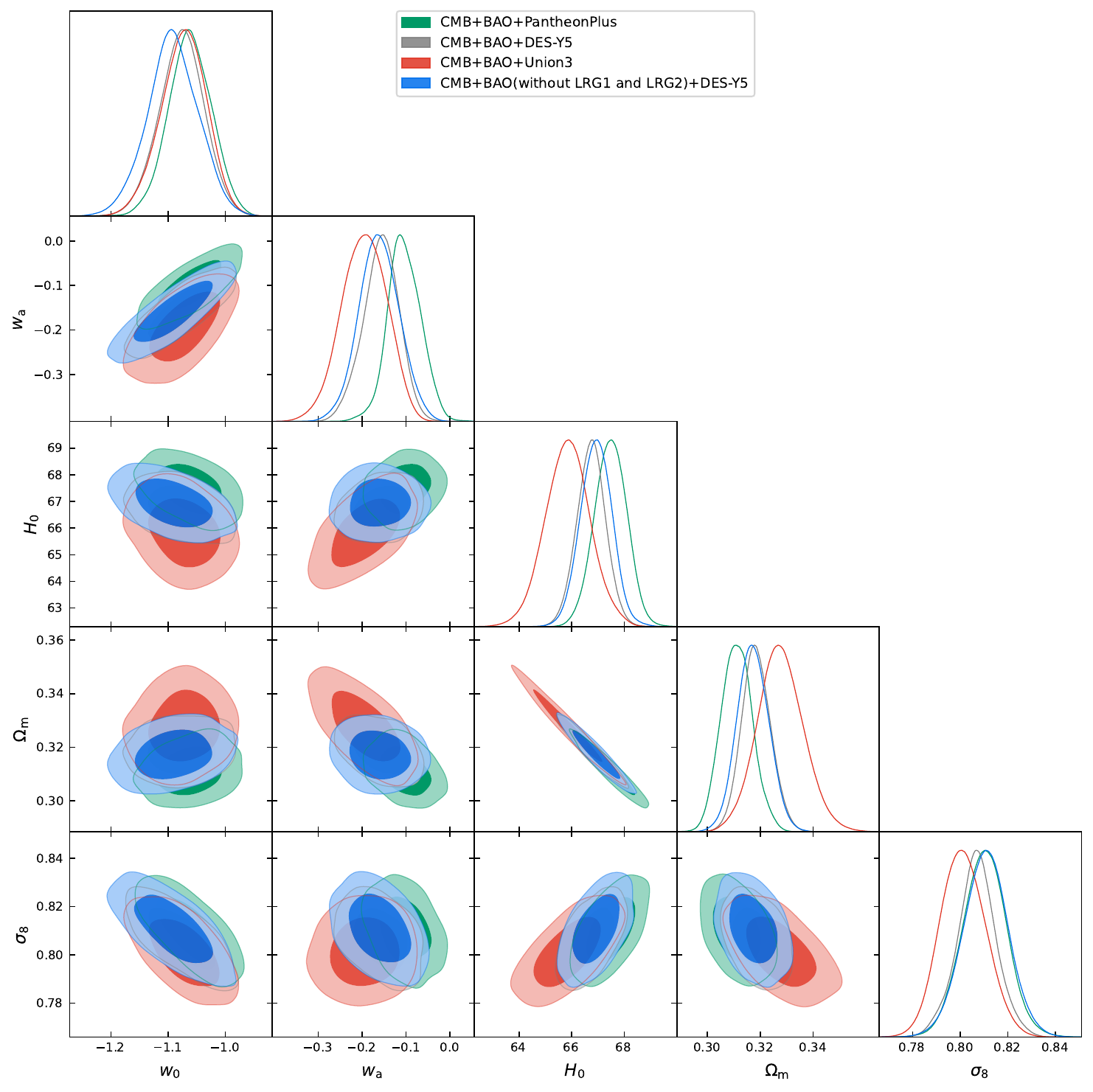}  
            \caption{EXP Parameterization}
            \label{subfig:i}
        \end{subfigure}
    \end{tabular}
    
    \caption{The triangular plots of the key parameters for the $\Lambda$CDM, $w$CDM, CPL, JBP, FSLL I, FSLL II, BA, LOG, EXP models derived from the dataset combinations of CMB+BAO+PantheonPlus, CMB+BAO+DES-Y5, CMB+ BAO+Union3 and CMB+BAO(without LRG1 and LRG2)+DES-Y5.}
    \label{FIG1}
\end{figure*}
\begin{figure*}[htbp]
    \centering
    \begin{tabular}{cc}  
        \begin{subfigure}[b]{0.47\textwidth}
            \centering
            \includegraphics[width=\linewidth]{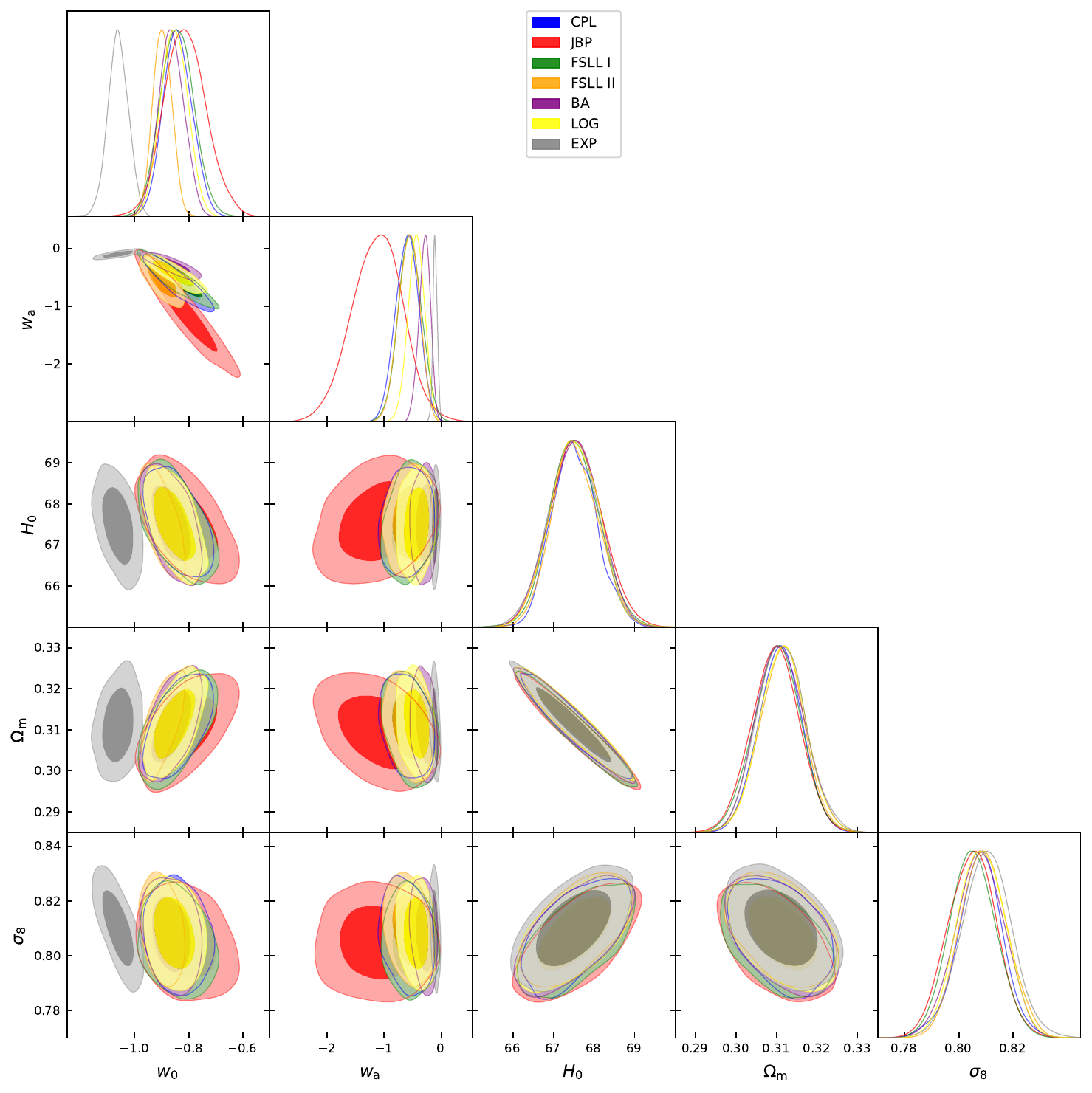}  
            \caption{CMB+BAO+PantheonPlus}  
            \label{PP}
        \end{subfigure}
        &
        \begin{subfigure}[b]{0.47\textwidth}
            \centering
            \includegraphics[width=\linewidth]{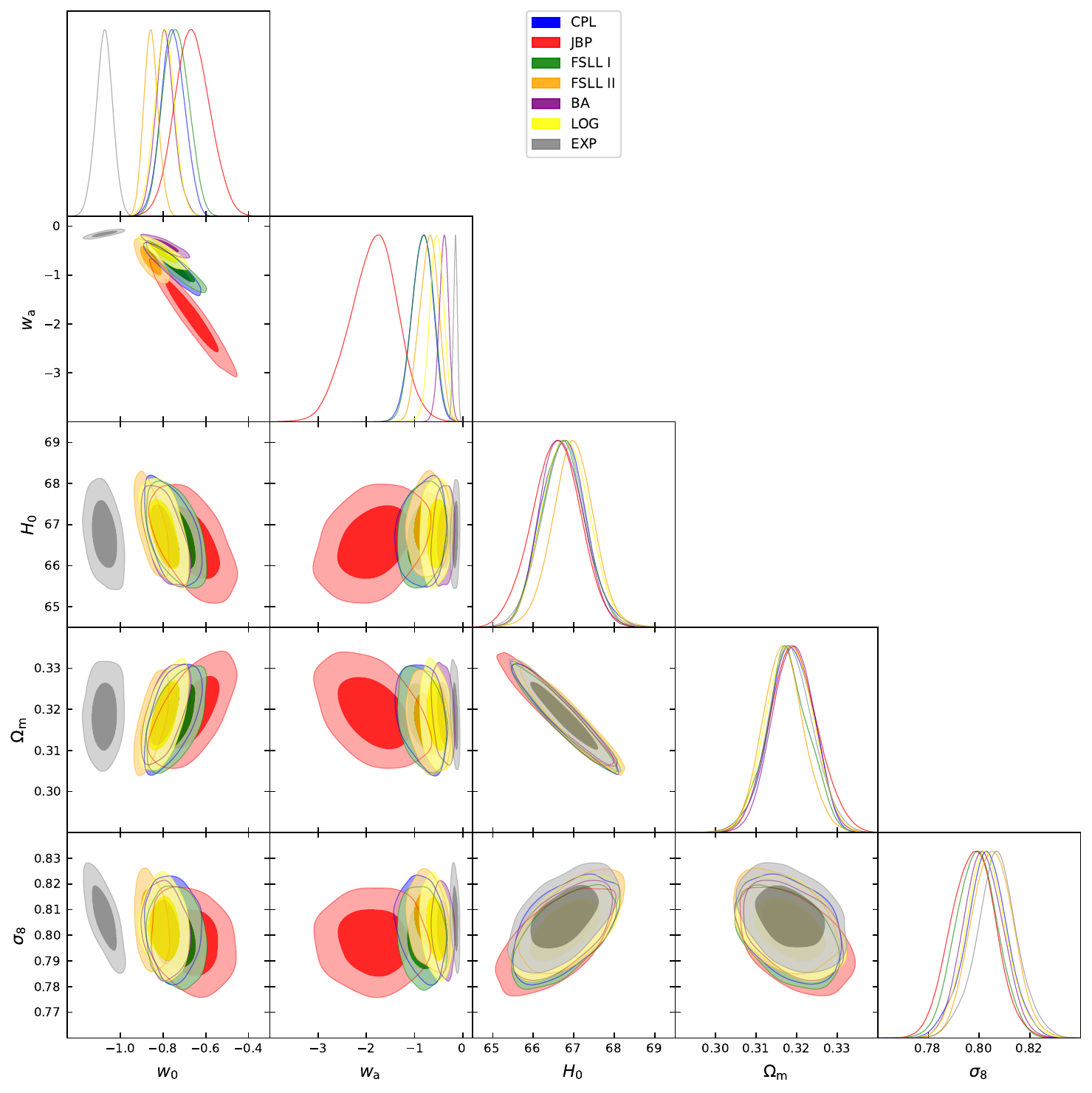}  
            \caption{CMB+BAO+DES-Y5}  
            \label{D5}
        \end{subfigure}
        \\[10pt]  
        
        \begin{subfigure}[b]{0.47\textwidth}
            \centering
            \includegraphics[width=\linewidth]{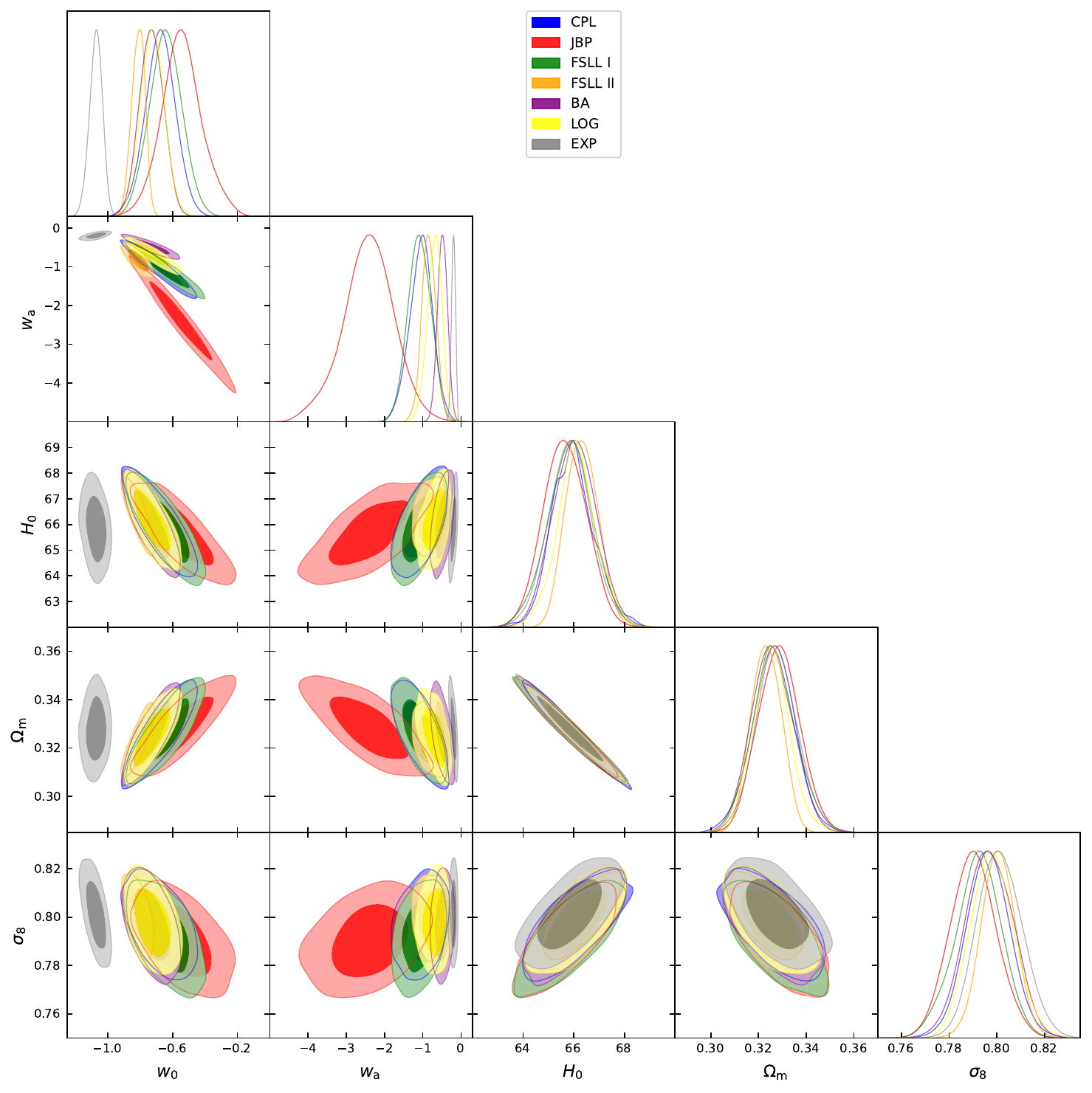}  
            \caption{CMB+BAO+Union3}  
            \label{U3}
        \end{subfigure}
        &
        \begin{subfigure}[b]{0.47\textwidth}
            \centering
            \includegraphics[width=\linewidth]{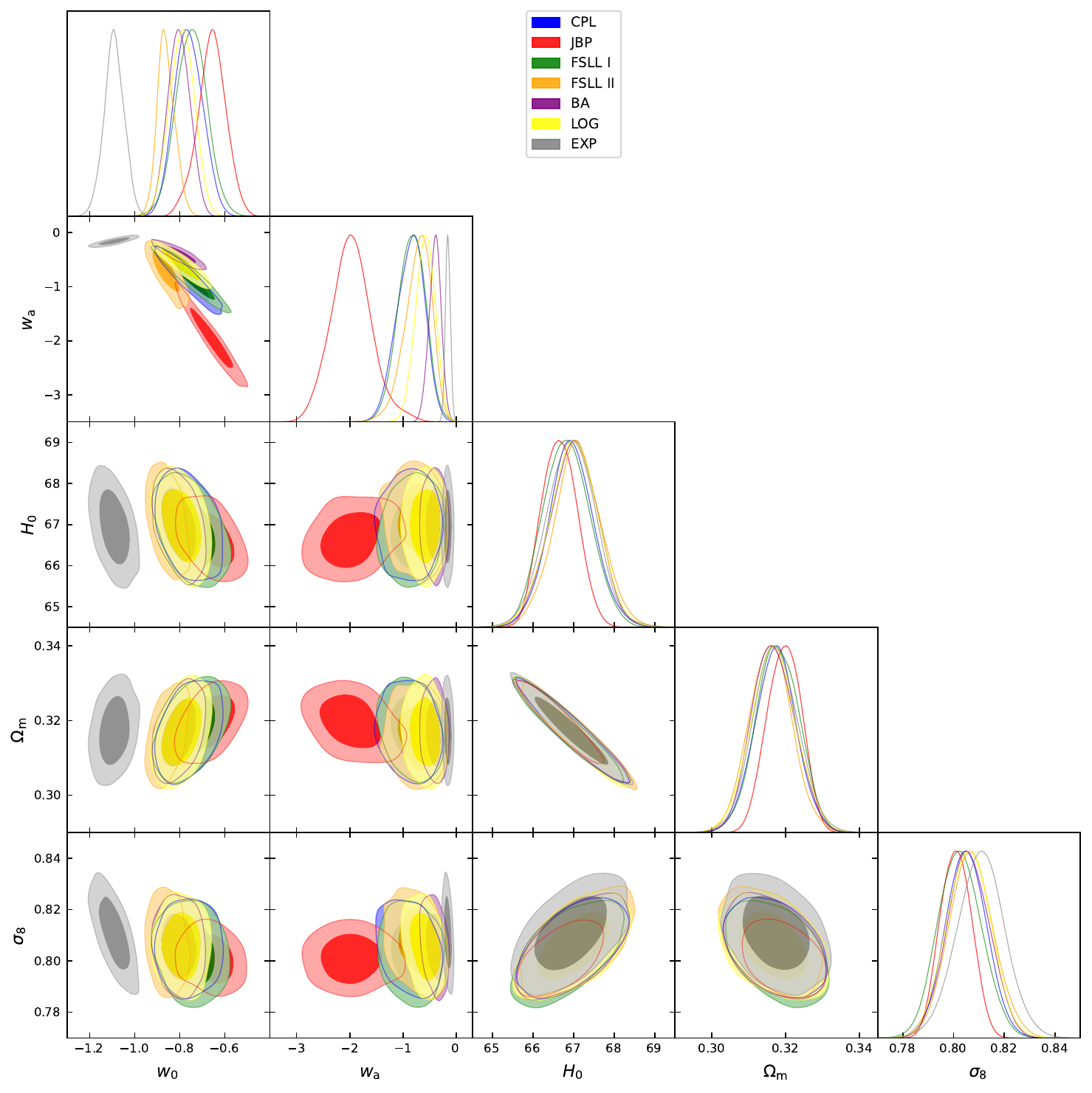}  
            \caption{CMB+BAO(without LRG1 and LRG2)+DES-Y5}  
            \label{LRG}
        \end{subfigure}
    \end{tabular}
    
    \caption{The triangular plots of the key parameters for each two-parameter DDE model under the CMB+BAO+ PantheonPlus, CMB+BAO+DES-Y5, CMB+BAO+Union3, and CMB+BAO (without LRG1 and LRG2)+DES-Y5 datasets.}
    \label{FIG2}  
\end{figure*}
\begin{figure*}[htbp]
    \centering
    \begin{tabular}{cc}  
        \begin{subfigure}[b]{0.47\textwidth}
            \centering
            \includegraphics[width=\linewidth]{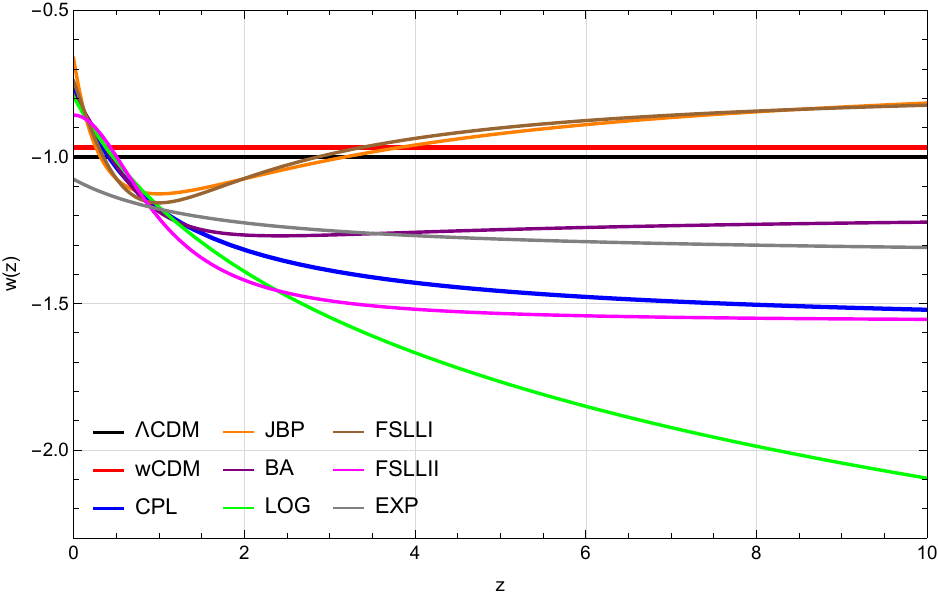}  
            \caption{$w(z)$}  
            \label{w(z)-D5}
        \end{subfigure}
        &
        \begin{subfigure}[b]{0.47\textwidth}
            \centering
            \includegraphics[width=\linewidth]{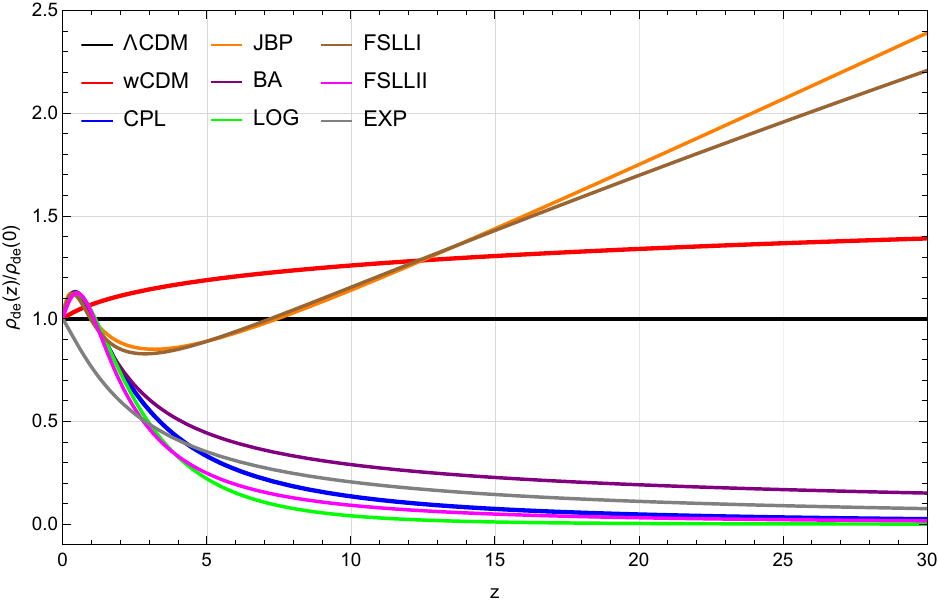}  
            \caption{${\rho }_{de}(z)/{\rho }_{de}(0)$}  
            \label{rho-D5}
        \end{subfigure}
    \end{tabular}
    
    \caption{$w(z)$ and ${\rho }_{de}(z)/{\rho }_{de}(0)$ as functions of redshift $z$ in CMB+BAO+DES-Y5. The parameter values are the central values of each model in Table~\ref{tb:result_combined}.}
    \label{FIG3}  
\end{figure*}
\begin{figure*}[htbp]
    \centering
    \begin{tabular}{cc}  
        \begin{subfigure}[b]{0.45\textwidth}
            \centering
            \includegraphics[width=\linewidth]{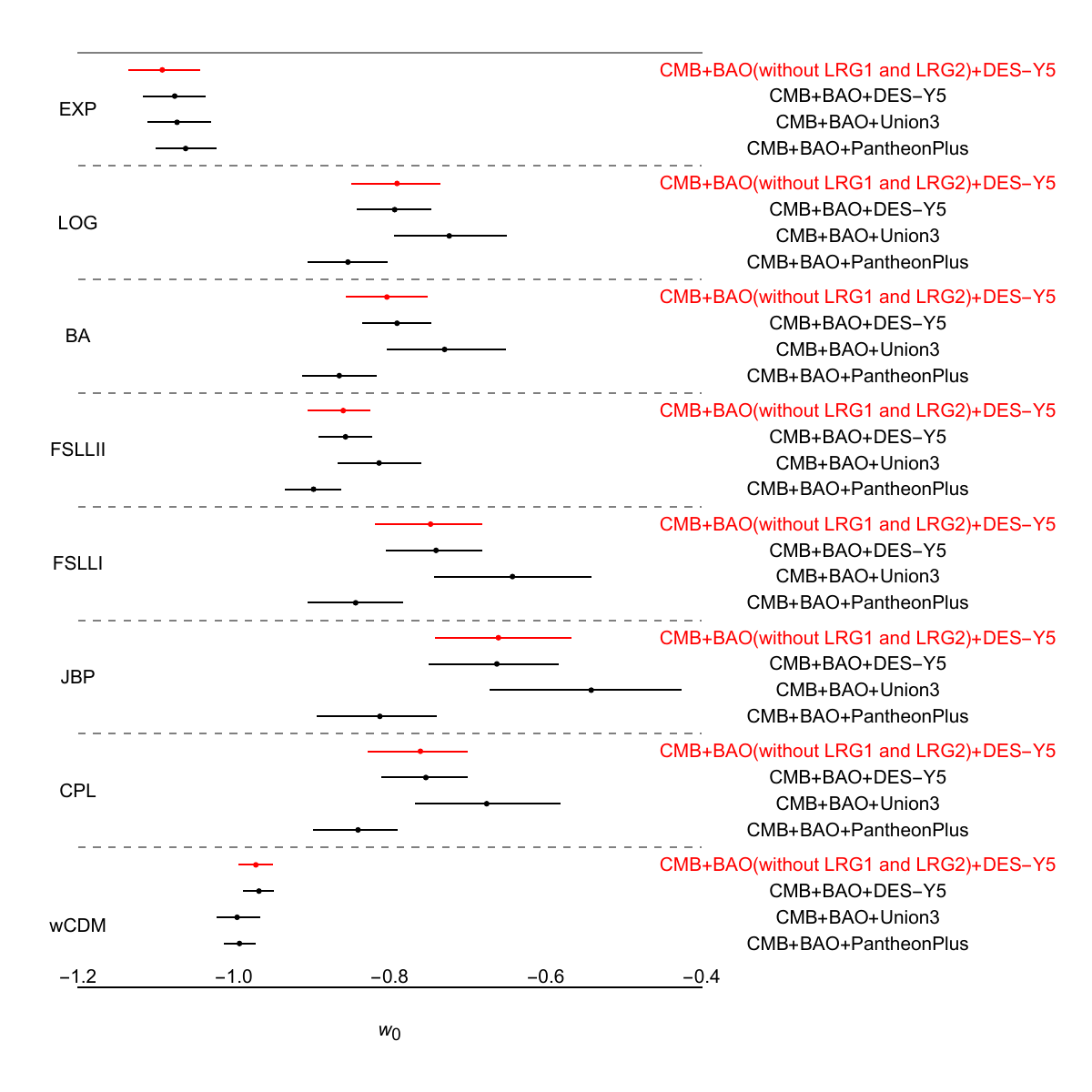}  
            \caption{$w_0$}  
            \label{w0}
        \end{subfigure}
        &
        \begin{subfigure}[b]{0.45\textwidth}
            \centering
            \includegraphics[width=\linewidth]{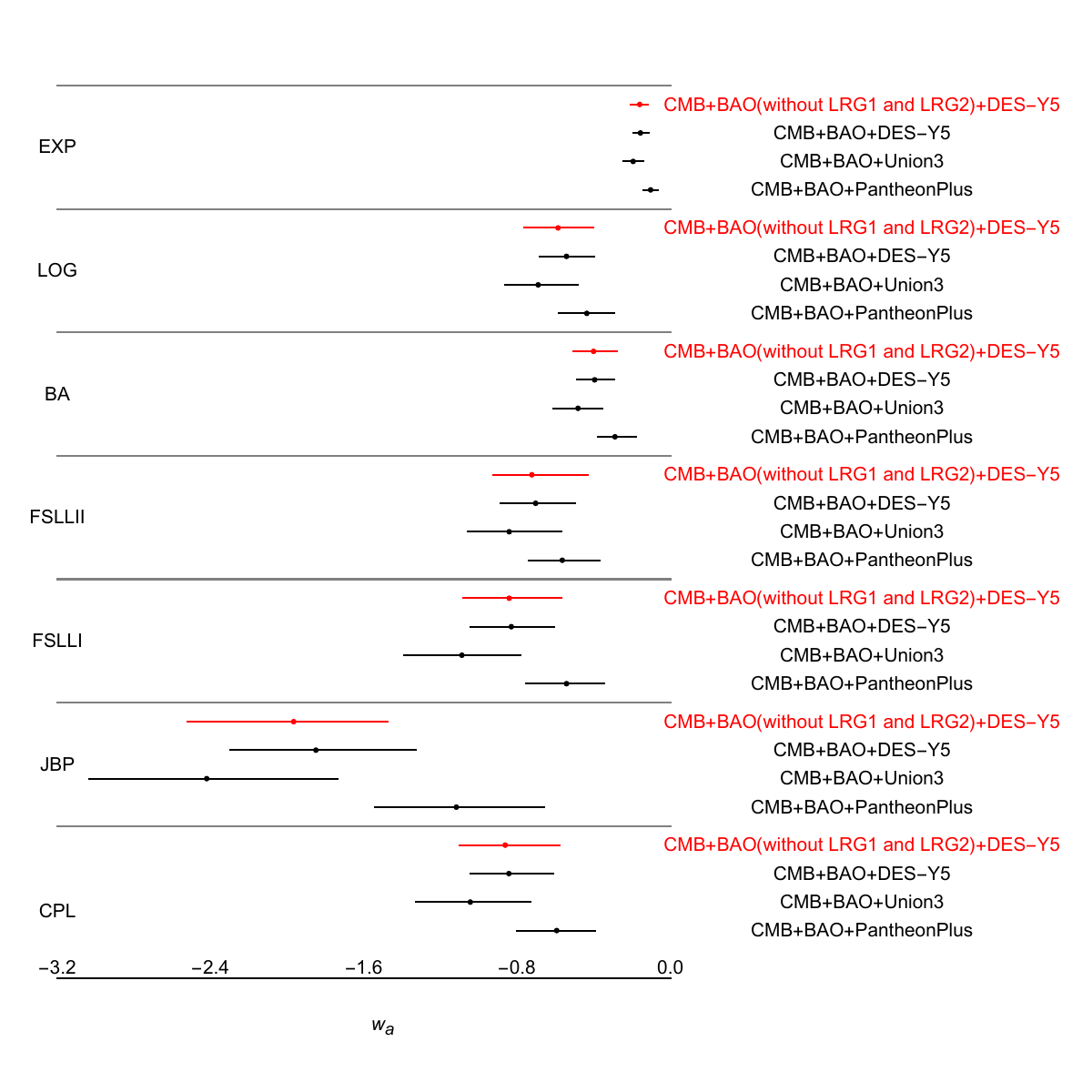}  
            \caption{$w_a$}  
            \label{wa}
        \end{subfigure}
        \\[10pt]  
        
        \begin{subfigure}[b]{0.45\textwidth}
            \centering
            \includegraphics[width=\linewidth]{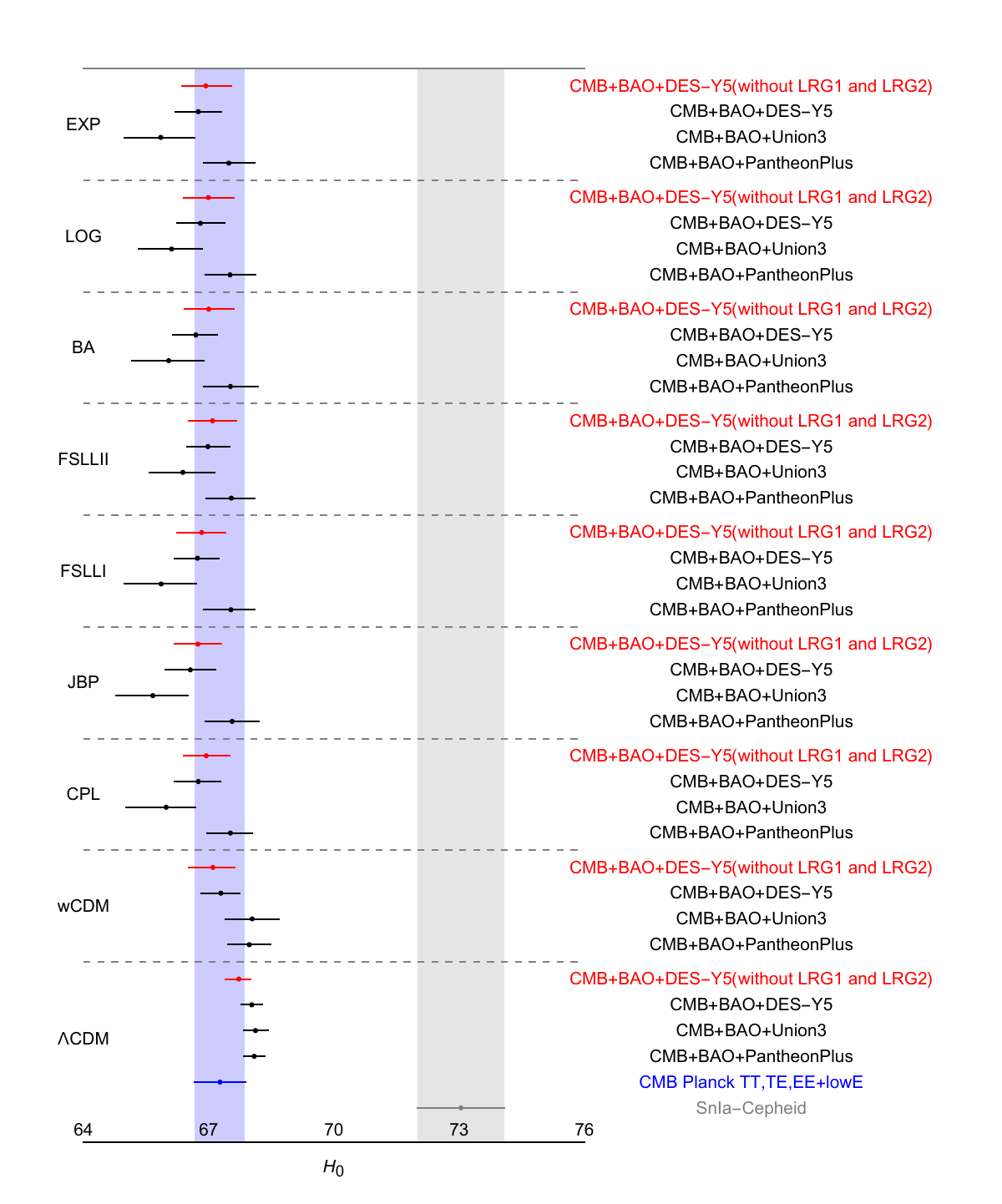}  
            \caption{$H_0$}  
            \label{H0}
        \end{subfigure}
        &
        \begin{subfigure}[b]{0.45\textwidth}
            \centering
            \includegraphics[width=\linewidth]{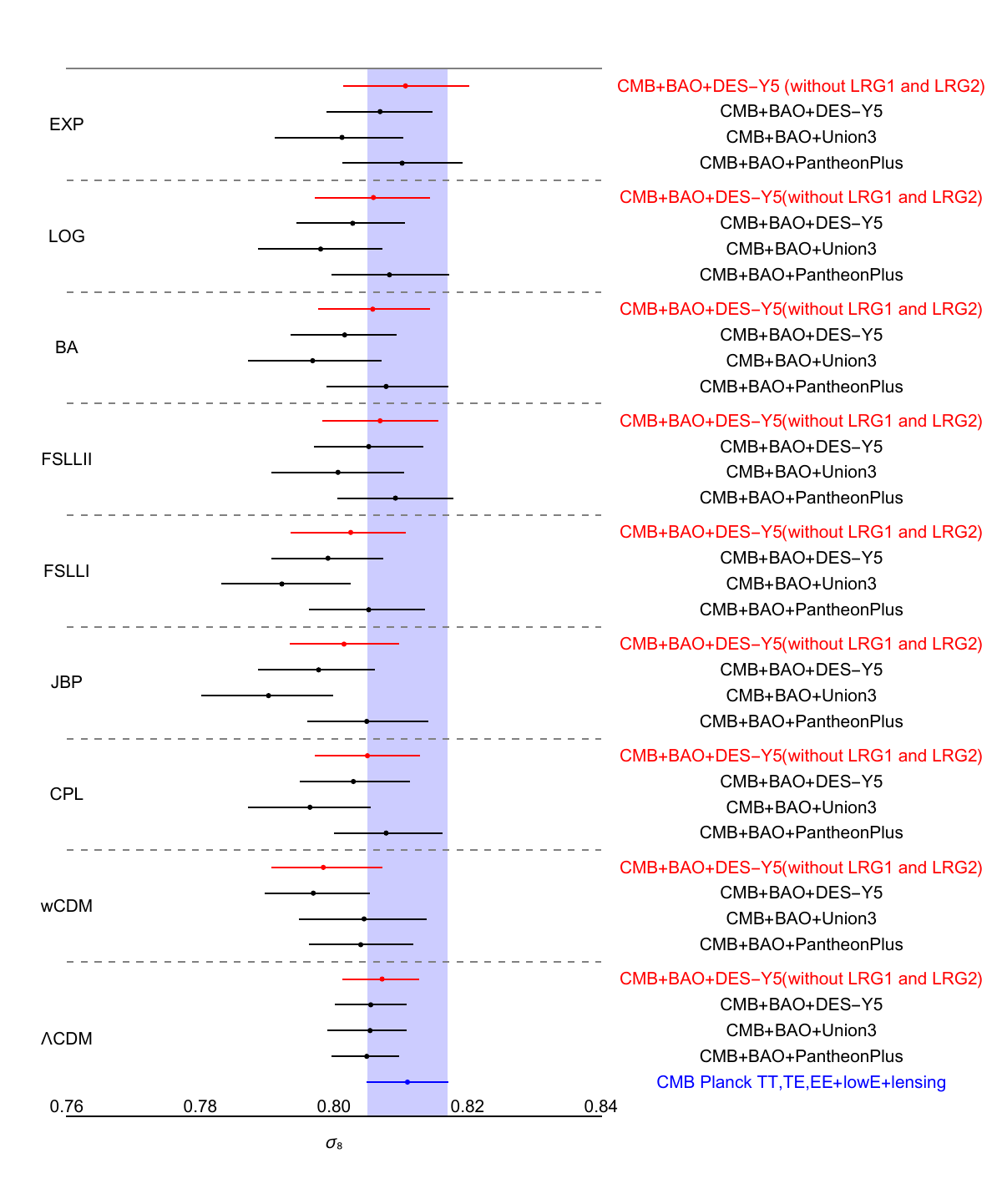}  
            \caption{$\sigma_8$}  
            \label{sigma8}
        \end{subfigure}
    \end{tabular}
    
    \caption{Marginal mean values of parameters $w_0$, $w_a$, $H_0$, and $\sigma_8$ at 68\% C.L. for all models. The blue band represents the constraints of Planck, and the gray band denotes the local constraints from SH0ES team.}
    \label{FIG4}  
\end{figure*}

\section{Discussion}
We will discuss from the following three perspectives:
\par Firstly, we discuss the changes in parameters when using different datasets under the same model. \\
1. In $w$CDM, the value of $w_0$ is comparable between CMB+BAO+PantheonPlus and CMB+BAO+Union3, but slightly larger in CMB+BAO+DES-Y5 and CMB+BAO(without LRG1 and LRG2)+DES-Y5. Except for EXP, $w_0$ and $w_a$ are inversely proportional for all other models, while only EXP exhibits approximately directly proportional characteristics. The uniqueness of EXP is further reflected in the insensitivity of its $w_0$ to the three SNIa compilations. The $w_0$ value of EXP decreases slightly only when LRG1 and LRG2 are removed. For other two-parameter DDE models, $w_0$ takes relatively larger values in CMB+BAO+Union3 and smaller values in CMB+BAO+PantheonPlus, while $w_a$ displays the opposite trend.
\\
2. Compared to CMB+BAO+DES-Y5, excluding LRG1 and LRG2 from BAO does not lead to a significant change in the constraints on the DE EOS parameters. The error bars are relatively large only when Union3 is adopted.
\\
3. In $\Lambda$CDM, the substitution of three SNIa data does not produce an obvious effect. When LRG1 and LRG2 are removed from BAO, $H_0$ increases and ${\Omega }_{m}$ decreases. For $w$CDM, using PantheonPlus or Union3 has little impact on the results. When DES-Y5 is used, $H_0$ decreases while ${\Omega }_{m}$ increases; this trend becomes more pronounced after excluding LRG1 and LRG2 from the BAO. Additionally, as shown in Figures~\ref{H0} and~\ref{sigma8}, compared to CMB+BAO+DES-Y5, the constraints of CMB+BAO(without LRG1 and LRG2)+DES-Y5 slightly shift the values of $H_0$ and ${\sigma }_{8}$ closer to the results derived from Planck 2018.
\\
4. $H_0$ and $\Omega_m$ exhibit a clear anti-correlation for all models, while $H_0$ shows a relatively weak positive correlation with $\sigma_8$. In all two-parameter DDE models, the values of $H_0$ and $\sigma_8$ are relatively lower with CMB+BAO+Union3 but higher with CMB+BAO+PantheonPlus. Moreover, the error bars of the three derived parameters are relatively large under CMB+BAO+Union3.
\\

\par Secondly, taking a different approach, we examine the trends in various parameters in different models using the same dataset. 
\\
1. Across all datasets, $w_0$ for EXP is obviously smaller, even falling below -1 at 68\% C.L.. In contrast, $w_0$ for JBP is slightly larger in all datasets, and its error bar is also the largest among all models. In Figure~\ref{w0}, we can see that the $w_0$ value for FSLL II is slightly lower. For other DDE models, $w_0$ is essentially stable with no obvious variations. Compared to $w$CDM, the two-parameter DDE models deviate significantly from $w_0=-1$. Among these models, only JBP exhibits a slightly lower degree of deviation: it only exceeds -1 at the 95\% C.L. under the CMB+BAO+PantheonPlus dataset (this can be attributed to its large error bar), while all other models (excluding EXP) exceed -1 at the 3$\sigma$ level.
\\
2. As shown in Figure~\ref{wa}, the values of $w_a$ derived from JBP are relatively small across all datasets, and the error bars are notably large. In contrast, EXP produces relatively higher $w_a$ values in all datasets, with the smallest error bars among all models and the strongest constraint power. The degree of deviation for $w_a$ from its standard value is even greater than that of $w_0$, and all models deviate from $w_a=0$ at 99\% C.L..
\\
3. It can be seen from Figures~\ref{H0} and~\ref{sigma8} that the three derived parameters of each model show little variation under CMB+BAO+PantheonPlus, indicating high consistency. Meanwhile, we find that when the parameter space is expanded, the marginal constraints of $H_0$ and $\sigma_8$ remain stable. In other datasets, the $H_0$ values of FSLL II are slightly higher. In general, compared to $\Lambda$CDM and $w$CDM, the two-parameter DDE models have lower values of $H_0$, which is not helpful in alleviating $H_0$ tension.
\\
4. Examining the impact of excluding LRG1 and LRG2 from BAO on the three derived parameters, we find that the parameters exhibit little variation in the two datasets. Only the $\Omega_m$ of JBP is slightly larger than that of other models in CMB+BAO(without LRG1 and LRG2)+DES-Y5; however, this difference becomes insignificant when using CMB+BAO+DES-Y5.
\\
5. In all datasets, $\sigma_8$ is slightly smaller for JBP and larger for EXP. As shown in Figure~\ref{sigma8}, in CMB+BAO+Union3, the $\sigma_8$ values of the two-parameter DDE models are slightly lower than those of $\Lambda$CDM and $w$CDM, which exerts a certain effect on alleviating $\sigma_8$ tension. In general, the variation of the three derived parameters is relatively small compared with that of the DE EOS parameters.
\\
\par Lastly, to compare the performance differences between various DDE models and the $\Lambda$CDM model, it is important to consider the additional parameters involved. The $w$CDM model has one additional degree of freedom ${w}_{0}$, while other DDE models are two-parameter models, each with two additional degrees of freedom, ${w}_{0}$ and ${w}_{a}$. In this work, we employ the AIC to appropriately penalize the additional parameters introduced by each model, and combine this with Bayesian evidence $\Delta\ln\mathcal{Z}$ to jointly quantify the goodness-of-fit for all models across different observational combinations. Notably, the judgments of AIC and $\Delta\ln\mathcal{Z}$ are similar in most cases but not completely consistent, so they will be discussed separately below.\\
1. When using the CMB+BAO+PantheonPlus dataset, all models except JBP have lower AIC values than $\Lambda$CDM, indicating a preference for these extensions. Among them, FSLL II achieves the lowest AIC, followed by LOG. However, the Bayesian evidence is different. All extended models yield $\Delta\ln\mathcal{Z} < 0$, indicating that none is statistically preferred over the $\Lambda$CDM model. Among them, LOG performs relatively best, with $\Delta\ln\mathcal{Z} = -0.51$.
\\
2. In CMB+BAO+DES-Y5, the AIC values of all DDE models are better than those of the $\Lambda$CDM model, with the LOG, BA, and FSLL II models being relatively more prominent. The Bayesian evidence basically follows this trend: the $\Delta\ln\mathcal{Z}$ value of the CPL is 4.55, and that of the LOG is 4.33, both of which yield strong evidence. The $\Delta\ln\mathcal{Z}$ values of the FSLL I, FSLL II and BA models are also greater than 3. All models except the $w$CDM model outperform the $\Lambda$CDM model.
\\
3. With CMB+BAO+Union3, all two-parameter DDE models produce lower AIC values than $\Lambda$CDM, reflecting a tendency to favor these models. The improvement is relatively pronounced for the LOG, EXP, and BA models, while the remaining models show only slight progress. Bayesian evidence also favors the LOG model the most, with a $\Delta\ln\mathcal{Z}$ value of 3.33, indicating strong evidential support. The $\Delta\ln\mathcal{Z}$ values of the FSLL II and CPL models are 2.23 and 2.06, respectively, which also receive moderate evidence. However, the other models do not show an obvious advantage over the $\Lambda$CDM model.
\\
4. Finally, we focus on the impact of LRG1 and LRG2 in BAO on model preference. Compared to CMB+BAO+DES-Y5, the preference for different models in CMB+BAO(without LRG1 and LRG2)+DES-Y5 changes significantly: the CPL, JBP and BA models have lower AIC values, which are favored to a certain extent. The same is true for Bayesian evidence. The $\Delta\ln\mathcal{Z}$ values of the JBP and CPL models are 2.76 and 2.00, respectively, providing moderate evidence. The other models show no obvious advantage over the $\Lambda$CDM model. In particular, the JBP model does not perform well in CMB+BAO+DES-Y5, but receives certain evidential support after excluding LRG1 and LRG2 from the BAO.
\\

\label{sec:sum}
\section{Summary}
\label{summary}
\par In this paper, we analyze the $\Lambda$CDM model and DDE parameterized models, placing strict constraints on key cosmological parameters using four datasets: CMB+BAO+PantheonPlus, CMB+BAO+DES-Y5, CMB+BAO+Union3 and CMB+BAO(without LRG1 and LRG2)+DES-Y5.

The study performs a comparative analysis based on the $\Lambda$CDM model, including the single-parameter DDE model: $w$CDM, and six two-parameter DDE models: CPL, JBP, FSLL, BA, LOG, and EXP. Our work focuses on the DE EOS parameters ${w}_{0}$ and ${w}_{a}$, while also constraining the important derived parameters ${H}_{0}$, ${\sigma }_{8}$ and ${\Omega }_{m}$. To ensure the timeliness and reliability of the analysis, our CMB data include Planck PR3 TTEE+lowE and the newly released Planck PR4 CamSpec high-$l$ TTTEEE. For the BAO data, DESI DR2 is also used. Finally, three SNIa datasets are combined: PantheonPlus, DES-Y5, and Union3.

Our analysis integrates the relatively scattered results in the field. Through a unified framework and standards, we compare the characteristics and advantages of different models and datasets, highlighting the systematicness and comprehensiveness of the research. 
Based on the above analysis, we find that LOG performs best in three different SNIa samples, and FSLL II is also relatively favored. After removing LRG1 and LRG2 from BAO, JBP and CPL obtain stronger evidence. Meanwhile, the two-parameter DDE models exhibit the best performance under CMB+BAO+DES-Y5. When combined with the $3\sigma$ level deviations of $w_0$ and $w_a$ from their standard values, our analysis provides the most robust evidence in favor of DDE.
Regarding Hubble tension, it is difficult to alleviate it only by adopting different parameterizations of the DDE models. However, CMB+BAO+PantheonPlus provides relatively high $H_0$ values, making it more suitable to investigate Hubble tension. In contrast, the $\sigma_8$ values in CMB+BAO+Union3 are relatively small, which helps mitigate the $\sigma_8$ tension. 
Furthermore, except for EXP, the two-parameter DDE models produce larger $w_0$ under CMB+BAO+Union3 and smaller $w_0$ under CMB+BAO+PantheonPlus, while $w_a$ exhibits the opposite trend. The JBP model has slightly larger $w_0$ values with the largest error bars among all models. 
Finally, compared to CMB+BAO+DES-Y5, the constraints of CMB+BAO(without LRG1 and LRG2)+DES-Y5 do not significantly alter the results of DE EOS, but slightly pull the values of $H_0$ and $\sigma_8$ toward the results of CMB.


In the end, our findings offer inspiration for further in-depth research. Specifically, they encourage an exploration of the physical mechanisms for high-performance parameterized models such as LOG and FSLL II. Furthermore, this research can help explore why the parameters of these models can better fit specific datasets. This may provide new theoretical clues for revealing the essence of dark energy.
\\

\vspace{5mm}
\noindent {\bf Acknowledgments}

We acknowledge the use of HPC Cluster of Tianhe II in National Supercomputing Center in Guangzhou. Lu Chen is supported by grants from NSFC (grant No. 12105164). This work has also received funding from project ZR2021QA021 supported by Shandong Provincial Natural Science Foundation and the Youth Innovation Team Plan of Colleges and Universities in Shandong Province (2023KJ350).



\bibliography{example} 
\end{document}